\pdfoutput=1
\documentclass[useAms,usenatbib]{mn2e}

\usepackage{graphicx}
\usepackage{lscape,graphicx}
\usepackage{amssymb}

\title{On time-dependent orbital complexity in gravitational N-body simulations}

\author[N. T. Faber, C. M. Boily and S. Portegies Zwart]
       {N. T. Faber$^{1,2,3}$\thanks{E-mail: nfaber@science.uva.nl
           (NF)}, C. M. Boily$^{1}$ and S. Portegies Zwart$^{2,3}$
         \\ $^{1}$Observatoire Astronomique, Universit\'e de
         Strasbourg and CNRS UMR 7550, 11 rue de l'Universit\'e, 67000
         Strasbourg, France
         \\ $^{2}$Sterrenkundig Instituut 'Anton
         Pannekoek', University of Amsterdam, Kruislaan 403, 1098 SJ
         Amsterdam, The Netherlands\\ $^{3}$Section Computational
         Science, University of Amsterdam, Kruislaan 403, 1098 SJ
         Amsterdam, The Netherlands}

\begin{document}

%Paragraph Status: 0=nothing done; 1=to be re-read; 2=revision finished

%\begin{frontmatter}



\newcommand{\hh}[1]{^{#1}}

\newcommand{\mybold}[1]{#1}


\maketitle
%\address{}

\begin{abstract}
We implement an efficient method to quantify time-dependent orbital
complexity in gravitational $N$-body simulations. The technique, which
we name DWaTIM, is based on a discrete wavelet transform of velocity
orbital time series. The wavelet power-spectrum is used to measure
trends in complexity continuously in time. We apply the method to the
test cases $N=3$ Pythagorean- and a perturbed $N=5$ Caledonian
configurations.  The method recovers the well-known time-dependent
complexity of the dynamics in these small-$N$ problems. We then apply
the technique to an equal-mass collisional $N=256$ body simulation ran
through core-collapse.  We find that a majority of stars evolve on
relatively complex orbits up to the time when the first hard binary
forms, whereas after core-collapse, less complex orbits are found on
the whole as a result of expanding mass shells.
\end{abstract}


%\end{frontmatter}

% Preliminary remark about the DWT plots: 
% Plots with pgsitf(0) for both the plot and the wedge: Figure 1: sinusoid_bfx
% pgsitf(2) for DWT_Plummer (a) both plot and wedge, pgsitf(1) for plot and pgsitf(2) for wedge of (b)
% pgsitf(1) for plot and pgsitf(2) for wedge for Caledonian plot
% Plots with pgsitf(2) for both the plot and the wedge: the rest

\section{Introduction}
\label{Sect:Introduction}

Computer simulations of the gravitational $N$-body problem aim to
solve the set of $3N$ second order ordinary differential equations
\begin{equation} \label{Eqn:grav}
  \bmath{F}_{i}=m_{i}\bmath{\ddot{r}}_{i}=-Gm_{i}\sum_{l=1,l\neq
    i}^{l=N}\frac{m_{l}(\bmath{r}_{i}-\bmath{r}_{l})}{|\bmath{r}_{i}-\bmath{r}_{l}|^{3}},\,
  (i=1,...,N).
\end{equation}
Here $G$ is the gravitational constant, $\bmath{F}_{i}$ and
$\bmath{\ddot{r}}_{i}$ denote the force and the acceleration exerted
by the $N-1$ particles $l$ on particle $i$ of mass $m_{i}$ at position
$\bmath{r}_{i}$. A system with $N>2$ is in general chaotic, i.e., the
solution to Eq.\,(\ref{Eqn:grav}) is known to be sensitive to the
initial conditions. Miller (1964) was the first to show that
individual orbits calculated from neighbouring initial configurations
diverge on a short time-scale proportional to $4\xi/N$, where $\xi$ is
the mean time between two subsequent close encounters of a
particle. Even when computed with double precision arithmetic, the
differences between the two integrations become comparable to the
characteristic length- and velocity-scales of the cluster within only
a few crossing times. Thus one may argue that the positions and
velocities obtained from an $N$-body simulation are a fair rendition
of the problem at hand in a statistical sense only
(\citealt{Aarseth75},
\citealt{Quinlan92}).  \citet{Goodman93} identify three mechanisms
responsible for the exponential growth of the distance between nearby
trajectories. Aside from numerical errors\footnote{A round-off error
at one time step can be considered as a change in the initial
conditions for the next time step.}, they point out that the
exponential instability of the solutions also arises through a
fluctuating mean gravitational field, or through inherently chaotic
orbits in a static field. However, for systems in dynamical
equilibrium and with near-spherical symmetry (such as e.g. globular
clusters), they conclude that the principal mechanism responsible for
chaos is the cumulative effect of near neighbour interactions,
i.e. two-body encounters. \\
\indent The exponential divergence of the $N$-body problem of
Eq.\,(\ref{Eqn:grav}) has been studied by several authors for systems
with up to $N=10^{5}$ particles (\citealt{Kandrup91};
\citealt{Kandrup01};
\citealt{Hemsendorf02}). The debate whether Miller's
instability is formally caused by chaos is still on-going (see e.g.,
\citealt{Kandrup03}; \citealt{Helmi07}). The assumed chaoticity of
such systems may be evaluated by computing the variational equations
associated to Eq.\,(\ref{Eqn:grav}) \citep{Miller71} and by retrieving
indicators of chaos such as, for instance, Lyapunov characteristic
numbers
\citep{Benettin76}. Other indicators of chaos commonly found in the literature are e.g., the
relative Lyapunov indicator \citep{Sandor00}, the mean exponential
divergence of nearby orbits \citep{Cincotta00} or the small-alignment
index \citep{Skokos01}. All these indicators accurately quantify the
exponential divergence of nearby trajectories.\\
\indent In this work, we propose to investigate time-dependent orbital
complexity in an $N$-body simulation. We define orbital complexity to
be a measure of the richness and non-triviality of the frequency
spectrum of an orbit at a given time $t$. The possible connection
between orbital chaos and orbital complexity has been pointed out by
\citet{Kandrup97}. In that work, the authors find a 
strong correlation between their measure of orbital complexity and
short time Lyapunov exponents. The notion of
complexity provides information about the orbital content of a
gravitational system governed by Eq.\,(\ref{Eqn:grav}) that is
complementary to the classical indicators of chaos mentioned
above. For instance, the concept of complexity is exploited by
\citet{Sideris02} to study the continuum limit in the case of
large-$N$ simulations. It is used to compare the orbital behavior
between particles evolving in smooth potentials and bodies orbiting in
the corresponding frozen $N$-body configurations. The notion of
complexity here allows to contrast the discreteness effects of the
$N$-body configuration to the orbital evolution obtained in the smooth
case. However, whereas a global measure of orbital complexity has been
implemented in several works (see e.g., Kandrup et al. 1997), a method
to measure the impact of instantaneous changes in orbital complexity
has not. A trajectory computed from Eq.\,(\ref{Eqn:grav}) can show
multiple, qualitatively distinct regimes in time. For instance, an
orbit may display arcs of relatively smooth motion such as e.g., an
unperturbed parabolic or hyperbolic orbit. At other times, the body
may be gravitationally bound in a binary system with a particle of
approximately the same mass. Likewise, the body may be temporarily
trapped in a complicated higher-order resonance, orbiting about a
massive central body. All these states of motion can be identified in
time by a suitable measure of complexity. The goal of this paper is
precisely to discuss such time-resolved complexity by introducing a
dedicated tool for complexity evaluation. The issue of formally
relating complexity to chaos will not be addressed here.\\
\indent Classical spectral methods and
Fourier transform based techniques (see \citealt{Laskar93};
\citealt{Carpintero98}; \citealt{Valluri98}) are of short execution
time and able to provide an accurate frequency domain representation
of a given orbit. A global measure of complexity can be retrieved by
such an approach. However, the Fourier transform suffers from the
problem of losing any time-dependent information on the motion of the
particle. In this paper, we present an accurate, easy-to-implement and
time-resolved complexity-detection tool for individual orbits in
$N$-body simulations. A state of dynamical equilibrium is assumed
throughout. We implement a specially-adapted method of time series
analysis that is based on a discrete wavelet transform. The method
post-processes the orbital data of a simulation and provides a
time-resolved measure of complexity. The present work mainly focuses
on a detailed description of the method and is structured as
follows. In Section 2 we introduce the identification technique for
time-dependent complexity. Section 3 presents applications to $N=2$,
$3$ and $5$-body problems. The analysis of a low-resolution $N=256$
equal-mass Plummer model is presented in Section $4$ and a brief
discussion of the results is given in Section $5$.

% checked!

\section{Method, Tests and Validation}
\label{Sect:method}

This section presents and tests a technique to quantify time-dependent
orbital complexity. In
\S\ref{Subsect:dwt}, we present the discrete wavelet transform
(hereafter DWaT) as an efficient tool to determine the base
frequencies of an orbit in a time-resolved manner. The indicator of
complexity obtained from the DWaT, namely the discrete wavelet
transform information measure (DWaTIM), is defined in
\S\ref{Subsect:complexity}.

\subsection{Wavelet transforms}
\label{Subsect:dwt}

\subsubsection{Definitions}
\label{Subsubsect:definitions}

% Intro
%\newline
\noindent The complexity 
of motion is analyzed by means of a wavelet transform (see also
\citealt{Gemmeke08}). The use of wavelet transforms has found applications in a
wide range of domains such as seismology, financial time series
processing or medical electrocardiogram studies; see \citet{Hubbard98}
for further references. A wavelet transform provides a time-frequency
representation of a time series $f(t_{q})$ by fitting a wavelet $\Psi$
to a set of points $\{t_{q}\}$. Whereas Fourier-based methods
decompose a signal into infinite sine and cosine functions,
effectively losing information at individual times $t_{q}$, the
wavelet transform offers precise localization in both the frequency-
and time-domain. Wavelet transforms remain band-limited however; they
are made up of not one but a limited range of several frequencies.

% The wavelet family
A wavelet family $\Psi_{a,\tau}$ is defined by the set of elemental
functions generated by scaling and translating a mother wavelet
$\Psi(t)$:
\begin{equation}
\Psi_{a,\tau}(t)= |a|^{-\frac{1}{2}}\Psi\left( \frac{t-\tau}{a}\right),
\end{equation}
where $a$ represents the scale variable and $\tau$
the translation variable ($a,\tau\in \mathbb{R}, a\neq 0$).

% The continuous wavelet transform
The continuous wavelet transform (CWT) is defined as the correlation
between a signal $S(t)\in L^{2}(\mathbb{R})$ (the space of square
summable functions) and the wavelet family $\Psi_{a,\tau}(t)$ for each
$a$ and $\tau$ (see e.g., \citealt{Daubechies92}):
\begin{equation} \label{Eqn:cwt} 
  \left\langle S(t),\Psi_{a,\tau}\right\rangle \equiv |a|^{-\frac{1}{2}}
  \int^{+\infty}_{-\infty}S(t) \bar{\Psi} \left(
  \frac{t-\tau}{a} \right) dt.
\end{equation}
Here $\langle S,\Psi_{a,\tau}\rangle$ denotes the wavelet
coefficients and $\bar{\Psi}$ is the complex conjugate of
$\Psi$. Equation (\ref{Eqn:cwt}) can be inverted to reconstruct the
original time series.  The CWT is known to produce a large amount of
wavelet coefficients which implies considerable CPU execution times
(see e.g., \citealt{Samar99}). In addition, the information the CWT
displays at closely spaced scales or at closely spaced time points is
highly correlated and thus unnecessarily redundant.

% The DWaT
For these reasons we instead compute a \textit{discrete} wavelet
transform (DWaT). The DWaT offers a highly efficient wavelet
representation that can be implemented with a simple recursive filter
scheme (\citealt{Mallat99}; \citealt{Daubechies92}). Unlike the
numerical CWT implementation which easily produces more than $10^{5}$
coefficients for a single orbital time series of $Q=8192$ data points,
the DWaT only produces as many coefficients as there are samples in
the time series, i.e. $Q$. This property of the DWaT of avoiding
redundant wavelet coefficients serves in defining a proper measure of
complexity, as we will show in \S\ref{Subsect:complexity}. For a given
choice of the mother wavelet function $\Psi(t)$ and for the discrete
set of parameters $a_{j}=2^{j}$ and $\tau_{j,k}=2^{j}k$ ($j$, $k$ $\in
\mathbb{Z}$), the wavelet family 
\begin{equation}
\Psi_{j,k}(t)= |a_{j}|\hh{-\frac{1}{2}}\Psi\left( \frac{t-\tau_{j,k}}{a_{j}}\right) 
=  2^{-\frac {j}{2}}\Psi({2^{-j}t-k})
\end{equation}
 defines an orthonormal basis of $L^{2}(\mathbb{R})$. The time series
 $f(t_{q})$ is sampled at $Q=2\hh{J}$ $(J\in\mathbb{N})$ constant time
 intervals of size $\Delta=t_{q+1}-t_{q}$. The discrete wavelet
 expansion then reads (see e.g., \citealt{Rosso06}, Equation [36])
\begin{equation}\label{Eqn:DWaT}
f(t)=\sum_{j=1}\hh{J}\sum_{k=1}\hh{2\hh{j-1}}\left\langle
f(t_{q}),\Psi_{j,k} \right\rangle_{\rm discrete}
\Psi_{j,k}(t)\Delta\hh{-1}.
\end{equation}
For simplicity we set $\Delta=1$ throughout
\S\ref{Sect:method}. Here $f(t)$ is the reconstructed signal and
$J=\log_{2}Q$ is the number of scales over which the time series
$f(t_{q})$ is analyzed. The DWaT coefficients $\langle
f(t_{q}),\Psi_{j,k}
\rangle_{\rm discrete}$ can be understood as a representation of the
wavelet power spectrum (or, energy) at scale $j$ and time $t_{q}$,
associated to the time series $f(t_{q})$. They represent the local
residual errors between successive signal approximations at scales $j$
and $j-1$. In what follows, we set
\begin{equation}
P_{j}(k)\equiv\left\langle f,\Psi_{j,k}\right\rangle_{\rm discrete}
\end{equation}
for a more convenient notation of the DWaT coefficients. As mentioned
earlier, there are $Q$ coefficients $P_{j}(k)$ and the number of
coefficients computed for resolution level $j$ is $2\hh{j-1}$. The
frequency band over which the $P_{j}(k)$ are computed is limited by
the frequency $1/(Q\Delta)$ (scale $j=1$) in the low-frequency domain
and by the Shannon-Nyquist critical frequency $f_{\rm cr}=1/(2\Delta)$ in
the high-frequency domain (scale $j=J$; see also \citealt[][\S 12.1
and \S 13.10]{Press02}). We refer to \citet{Samar99} for further
details about the wavelet representation.

% The mother wavelet
We use bi-orthogonal cubic spline functions as mother wavelets
(\citealt[][case $\{N,\tilde{N}\}=\{3,9\}$ in their Table 6.1]{Cohen92}),
\begin{equation}
\Psi(t)=\sum_{l}g_{l}\Theta_{3}(2t-l),
\end{equation}
where the $g_{l}$'s are known as basic spline coefficients and
\begin{equation}
\Theta_{3}(t)= \left\{ 
\begin{array}{lr}
(t+1)^{2}/2, & -1\leq t\leq 0\\
-(t-1/2)^{2}+3/4, & 0\leq t\leq 1\\
(t-2)^{2}/2, & 1\leq t\leq 2\\
0, & \rm{otherwise.}
\end{array}
\right.
\end{equation}
 This choice is motivated by three arguments. First and most
 importantly, spline functions provide an excellent time-frequency
 localization when compared to other mother wavelet candidates
 (\citealt{Ahuja05};
\citealt{Unser99}). Instantaneous changes in the dynamics are
accurately singled out by the DWaT. We further stress this point
in \S\ref{Subsect:toy_model}. Second, the use
of splines is computationally inexpensive
\citep{Thevenaz00} and provides further desirable
properties such as e.g., compact support and smoothness. (For an
exhaustive discussion on spline interpolation, see \citealt{Unser99},
\citealt{Thevenaz00}.) Finally, the use of a bi-orthogonal spline
mother wavelet also implies reduced border effects, an undesired
artifact of the wavelet transform algorithm (see
\S\ref{Subsubsect:bfx} below).

% The time series we use
We analyze the velocity time series
\begin{equation}
f(t_{q})= v_{\alpha_{i}}(t_{q})\qquad\qquad\qquad(\alpha=x,y,z),
\end{equation}
i.e. the $v_{x_{i}}$, $v_{y_{i}}$ and $v_{z_{i}}$ velocity components
of each particle $i$ $(i=1,...,N)$. We do not use the information
available on the positions of the bodies since there may be important
differences in magnitude between the beginning and the end of these
time series. Positional information is then likely to produce
pronounced DWaT border effects (see \S\ref{Subsubsect:bfx}).

% limit case of only 4 oscillations, the area affected by BFX is indeed 25%. This is
% also easily seen in the figure sinusoid_bfx shown here below.

%%%%%%%%%%%%%%%%%%%%%%%%%%%%%%%%%%%%%%%%%%%%%%%%%%%%%%%%%%%%%%%%%%%%%%%%%%%%%

% Figure 1: Sinusoid
%%%%%%%%%%%%%%%%%%%%%%

\begin{figure}
\centering
\includegraphics[height=130mm,width=80mm]{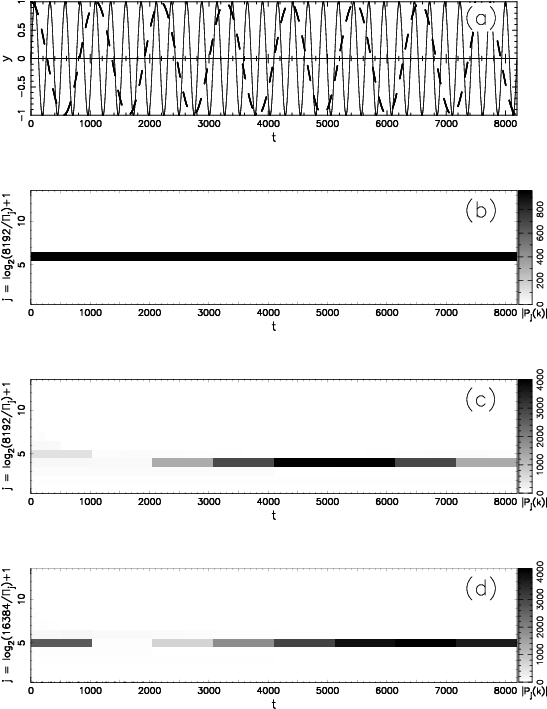}
\caption{Border effects of the discrete wavelet transform (DWaT)
of the two sinusoids $y_{1}$ and $y_{2}$ defined in
Eq.\,\ref{Eqn:sinusoid_bfx}. From top to bottom: (a) time series of
$y_{1}$ and $y_{2}$ (the dashed line is $y_{2}$), (b) DWaT of $y_{1}$, (c) DWaT
of $y_{2}$ with full border effects, (d) DWaT of $y_{2}$ with reduced
border effects.}
\label{Fig:sinusoid_bfx}
\end{figure}

%Caption Status: 2

\subsubsection{Border effects}
\label{Subsubsect:bfx}

%\newline
\noindent Border effects are an artifact of the
wavelet transform algorithm which enforces cyclical boundary
conditions on the data vector (\citealt{LoPresti96}; \citealt[][\S
13.10]{Press02}). Such border effects depend on the values held by the
two end points of the data vector, and may become important when both
ends of the data set differ greatly. In the following, we aim to
quantify the extent to which the diagnostics become erroneous due to
these edge effects.\\
\indent Figure \ref{Fig:sinusoid_bfx}(a) shows the two sinusoids
\begin{eqnarray} \label{Eqn:sinusoid_bfx}
y_{1} & = & \cos \left(2\pi \times \frac{32}{8192} q + \frac{3\pi}{2}
\right)\nonumber, \\ y_{2} & = & \cos \left(2\pi \times \frac{15/2}{8192} q\right),
\end{eqnarray}
sampled at $Q=8192$ intervals $(q=1,...,Q)$. The integration time
is commensurate to the periodicity of signal $y_{1}$. This is not the
case for signal $y_{2}$. Figure
\ref{Fig:sinusoid_bfx}(b) and (c) show their respective DWaT scalograms.  The
scalogram is a grey-shaded representation of the DWaT
coefficients $P_{j}(k)$. The darkest shade is for the largest values
$|P_{j}(k)|$; white means $P_{j}(k)=0$.  For both sinusoids a maximum
intensity is obtained for the scale that represents the base period
$\Pi_{j}$ of the respective signal. These are scales $j=6$ at period
$\Pi_{j}=2\hh{1-j}\cdot Q\Delta=2^{-5}\cdot 8192=256$ for $y_{1}$
and $j=4$ for $y_{2}$. The region in which the discrete wavelet
transform of $y_{2}$ suffers from border effects is known as the cone
of influence (\citealt{Moortel04}). The cone of influence is clearly
visible at both edges on Fig.\,\ref{Fig:sinusoid_bfx}(c). On the
left-hand edge of Fig.\,\ref{Fig:sinusoid_bfx}(c), for example, the
DWaT gives an artificially high excited mode at scale $j=5$ up to
$t=1024$. The same artifact is also found at scales of higher
frequencies, although the magnitude of the effect then diminishes and
is barely visible on the left-hand side of
Fig.\,\ref{Fig:sinusoid_bfx}(c). In addition, scale $j=4$ is wrongly
excited in the intervals $0\leq t\lesssim 3072$ and $7168\lesssim t<
8192$. The size of the cone of influence depends on the choice of the
mother wavelet and is especially significant when the spectrum of the
signal contains low frequencies, when the period compares to the
overall duration of the time series (Moortel et al. 2004).\\
\indent Figure \ref{Fig:sinusoid_bfx}(d) shows a situation where the border
effects have been reduced by extending the integration time to twice
the original interval\footnote{Alternately, the $y_{2}$ data vector
may also be padded with zeros until twice the original time interval
is reached.}. Doubling the time interval of analysis implies an
increased ratio between the signal length and its periodicity. This
allows for a higher resolution in the frequency domain (Carmona et
al. 1998). Furthermore, a larger portion of the DWaT remains
unaffected by the border discontinuity and thus a larger amount of
reliable information can be retrieved. In particular, the left-hand
and the right-hand border effects are diminished on
Fig.\,\ref{Fig:sinusoid_bfx}(d): for instance, the wrongly excited
mode at $j=5$ between $0\leq t\leq 1024$ in
Fig.\,\ref{Fig:sinusoid_bfx}(c) has been reduced by $\approx 82\%$ and
the scale $j=4$ artifacts of the right-hand side of that figure have
vanished completely. In the remainder of this paper, we apply the
technique of time series extension whenever the edge effects appear
significant. To take into account the remaining artifacts at the
beginning of the time series, we furthermore analyze the first
$\approx 2048$ data points (equivalent to the first $\approx 25\%$ of
the signal) of the DWaT analysis with particular caution. Although
Fig.\,\ref{Fig:sinusoid_bfx}(d) indicates that border effects may
influence more than the first 25\% of the $y_{2}$ analysis, we found
that this is in general not the case for signals produced by $N$-body
orbits (see
\S\ref{Subsubsect:sinusoid_complexity} and
\S\ref{Sect:few-body}). In what follows, we refer to these 25\% of the
analysis as potentially biased due to border effects.

% This choice of 25% is further motivated by the fact that we do not expect to analyze
% time series that show less than 4 dynamical times (or complete oscillations). In this

% Conclusion: We have been capable to reduce bfx i such a way that no SCALE has been excited at a significant level.
% This shows that at least a minimum 75% of the signal the magnitude is still to be discussed

\subsection{Measures of complexity}
\label{Subsect:complexity}

%\newline
\noindent  
Our goal is to obtain from the DWaT a quantitative estimate of the
time-dependent complexity of the frequency spectrum of a
trajectory. Exploiting some notions of information theory, we here
present the discrete wavelet transform information measure (DWaTIM) as
an efficient indicator for complexity. In what follows, we provide a
succinct overview of the concept. We follow closely the approach of
\citet{Rosso06} and
\citet{Martin06}. We refer the reader to these works for a more extended
discussion.

\subsubsection{Discrete wavelet transform information measure (DWaTIM)}
\label{Subsubsect:dwatim}

For a chosen time window of size $\kappa \Delta$ \mybold{(where $\kappa$ is an
arbitrary integer)} we compute the wavelet energy at each resolution
level $j$,
\begin{equation}
E_{j}=\sum_{k=1}\hh{2\hh{j-1}}|P_{j}(k)|^{2},
\end{equation}
and the total wavelet energy,
\begin{equation}
E=\sum_{j}E_{j},
\end{equation}
to obtain the so-called relative wavelet energy $p_{j}=E_{j}/E$ at
scale $j$. The DWaT then provides a probability distribution
\begin{equation}\label{Eqn:proba}
\mathcal{P}=\{p_{j}\},\quad (j=1,...,J),
\end{equation}
which weighs the base frequency $j$ in the reconstruction of the
original signal $v_{\alpha_{i}}(t_{q})$. By definition we have
$\sum_{j=1}^{J}p_{j}=~1$. \\
\indent The amount of disorder present at time $t$ in an orbital time series
can be quantified by determining the information needed to describe
the orbit at that time \citep{Shannon48}. An information measure
(hereafter IM) can be seen as a quantity that describes the
characteristics of the time-scale probability distribution
$\mathcal{P}$ of Eq.\,(\ref{Eqn:proba})
\citep{Rosso06}. The IM gives the amount of
information required per time unit to specify the state of the system
up to a given accuracy. The IM we use in this work is the Shannon
entropy (\citealt{Shannon48}; \citealt{Quiroga99}; \citealt{Sello03})
\begin{equation} \label{eqn:Shannon}
W_{S}[\mathcal{P}]=-K \sum_{j}p_{j}\log_{2}(p_{j}),
\end{equation}
where $K=1$ is an arbitrary numerical constant. A minimum of
information entropy $\min{\left(W_{S}[\mathcal{P}]\right)}=0$ is
obtained if $p_{j}=1$ for some scale $j$ and $0$ for all the remaining
scales. This situation only occurs for the ordered dynamics of a
periodic orbit with a single base frequency. Likewise, the state of
highest complexity is obtained for the case of a white noise
signal. For such an orbit, the entire band of base frequencies is
sampled by the DWaT in equal proportions, and $\mathcal{P}$ is
characterized by the uniform probability distribution
\begin{equation}\label{Eqn:proba_u}
\mathcal{P}_u=\left\{
\frac{1}{J},...,\frac{1}{J}\right\}.
\end{equation} 
In this case the IM is
$\max{\left(W_{S}[\mathcal{P}]\right)}=W_{S}[\mathcal{P}_{u}]=\log_{2}J$. For
a given velocity component $v_{\alpha_{i}}$
$(\alpha=x,y,z;\;i=1,...,N)$ we define the DWaTIM to be the measure
\begin{equation}\label{DWaTIM}
H_{\alpha_{i}}\equiv\left(\frac{W_{S}[\mathcal{P}]}{W_{S}[\mathcal{P}_{u}]}\right)_{\alpha_{i}},
\end{equation}
i.e. the Shannon entropy for velocity component $\alpha$ of particle
$i$ \mybold{normalized to the interval $[0,1]$}. The generalized
DWaTIM of particle $i$ is then obtained by taking the arithmetic mean
over the $3$ components,
\begin{equation}
\Upsilon_{i}\equiv\frac{H_{x_{i}}+H_{y_{i}}+H_{z_{i}}}{3}.
\end{equation}
Finally, the overall DWaTIM $\Upsilon_{\rm tot}$ of the system is computed
by averaging over all particles $i$,
\begin{equation}
\Upsilon_{\rm tot}\equiv\frac{\sum_{i=1}^{N}\Upsilon_{i}}{N}.
\end{equation}
In what follows we will argue that an increase in DWaTIM correlates
with an increase in the complexity of the underlying orbital dynamics
at that instant.

\subsubsection{Complexity of a sinusoid}
\label{Subsubsect:sinusoid_complexity}

%%%%%%%%%%%%%%%%%%%%%%%%%%%%%%%%%%%%%%%%%%%%%%%%%%%%%%%%%%%%%%%%%%%%%%%%%%%%%

% Figure 2: DWaTIM of sinusoid
%%%%%%%%%%%%%%%%%%%%%%%%%%%%%
\begin{figure}
\centering
\includegraphics[height=50mm,width=65mm]{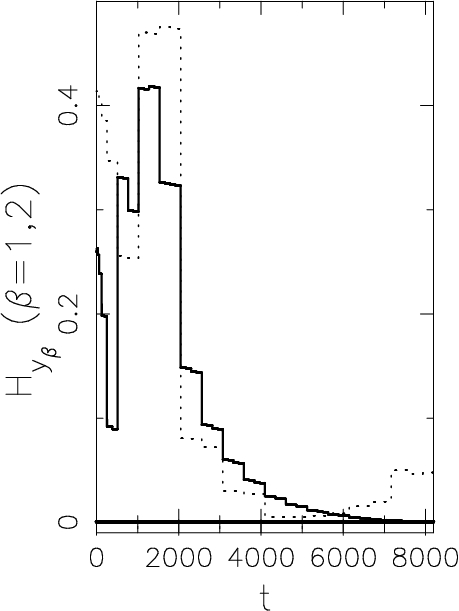}
\caption{Complexity measure for the sinusoids $y_{1}$ and $y_{2}$ of Fig.\,\ref{Fig:sinusoid_bfx}. 
The thick solid line confounded with the $x$-axis shows the DWaTIM
$H_{y_{1}}$ as obtained for $y_{1}$ from
Fig.\,\ref{Fig:sinusoid_bfx}(b). The upper solid line shows
the DWaTIM $H_{y_{2}}$ obtained for $y_{2}$ from
Fig.\,\ref{Fig:sinusoid_bfx}(d) in which border effects where reduced
by time series extension. The \mybold{dotted} line indicates the $H_{y_{2}}$
curve obtained for the case where border effects have not been treated
(see Fig.\,\ref{Fig:sinusoid_bfx}[c]).}
\label{Fig:sinusoid_dwatim}
\end{figure}

%Caption Status: 2
%%%%%%%%%%%%%%%%%%%%%%%%%%%%%%%%%%%%%%%%%%%%%%%%%%%%%%%%%%%%%%%%%%%%%%%%%%%%%

The temporal evolution of the DWaTIM indicator is computed by
subdividing the input signal in non-overlapping time windows of size
$\kappa \Delta$. We use $\kappa=2$. Figure
\ref{Fig:sinusoid_dwatim} shows our complexity analysis for the 
sinusoids $y_{1}$ and $y_{2}$ of \S\ref{Subsubsect:bfx}. The DWaTIM of
$y_{1}$, $H_{y_{1}}$\footnote{For a more convenient notation, we
denote the DWaTIMs obtained for the test cases $y_{\beta}$
$(\beta=1,2,I,II)$ presented in section \S\ref{Sect:method} by
$H_{y_{\beta}}$.}, is shown by the lower thick solid line. As expected
for a stationary signal with a unique base frequency, $H_{y_{1}}\simeq
6.13\cdot 10\hh{-4}$; the line almost coincides with the $y=0$ axis. In
what follows, this value can be considered as indicating zero
complexity.

We now discuss the pathological case of signal $y_{2}$ already mentioned in
\S\ref{Subsubsect:bfx}. The accuracy of the DWaTIM
depends on the frequency of the sinusoid. The DWaT has a better
resolution at high frequencies \citep{Samar99}. Conversely, it is more
difficult to obtain reliable measures of complexity for signals of low
frequencies. In addition, border effects are the most important for
such time series (see \S\ref{Subsubsect:bfx}). The time series
$y_{2}(t_{q})$ is a long-period, non-cyclical, unique-frequency
signal. The DWaTIM $H_{y_{2}}$ is shown by the upper solid line on
Fig.\,\ref{Fig:sinusoid_dwatim}. The line depicts the DWaTIM obtained
from the DWaT of Fig.\,\ref{Fig:sinusoid_bfx}(d), i.e. for the case
where border effects have been reduced by time series extension. (Note
that the border effects for that case still influence the result at
all times.) For the first 25\% of the diagnostic the effects are
dramatic: we find complexity values of up to $\approx 0.4$, where
$\approx 0$ is expected. The curve $H_{y_{2}}$ approaches zero for
$t\gtrsim 6500$ only. Our method of border effect reduction thus
brings little improvement to the quality of the diagnostic for that
signal (compared with no corrections at all, cf. \mybold{dotted} line
on Fig.\,\ref{Fig:sinusoid_dwatim} obtained from
Fig.\,\ref{Fig:sinusoid_bfx}[c]). However, one should bear in mind
that the DWaT approach stems from the idea of performing a
\mybold{\textit{multi}-frequency} analysis of signals that are well-resolved over the
time interval of interest. It is \mybold{difficult} to analyze
\mybold{with the DWaT} non-cyclic single-frequency signals of long period
\mybold{such as e.g., signal $y_{II}$} (\mybold{see also}
\citealt{Moortel04}; \citealt{Samar99}). \mybold{Our analysis of such 
sinusoids exposed the limits of the DWaTIM method; we here obtained
complexity results that were widely biased by border effects.}
\mybold{However}, in this work we aim to discuss the more generally
encountered multi-frequency \mybold{signals describing} orbital motion
in the gravitational $N$-body problem. For this type of signal, we
found that the complexity analysis yields \mybold{in general}
consistent results whenever
\mybold{the lowest frequency component of} the time series is resolved
over at least $\approx 8$ complete oscillations. Border effects then
play a minor role. \mybold{(For a further discussion of this, see the
results obtained for the multi-frequency signals of
\S\ref{Subsect:toy_model},
\S\ref{Subsect:binaries} or}
\S\ref{Subsect:pyth}.)

\subsection{Toy Models}
\label{Subsect:toy_model}

%%%%%%%%%%%%%%%%%%%%%%%%%%%%%%%%%%%%%%%%%%%%%%%%%%%%%%%%%%%%%%%%%%%%%%%%%%%%%

% Figure 3: Toy Model1
%%%%%%%%%%%%%%%%%%%%%%%%%%%%%

\begin{figure}
\centering
\includegraphics[height=100mm,width=80mm]{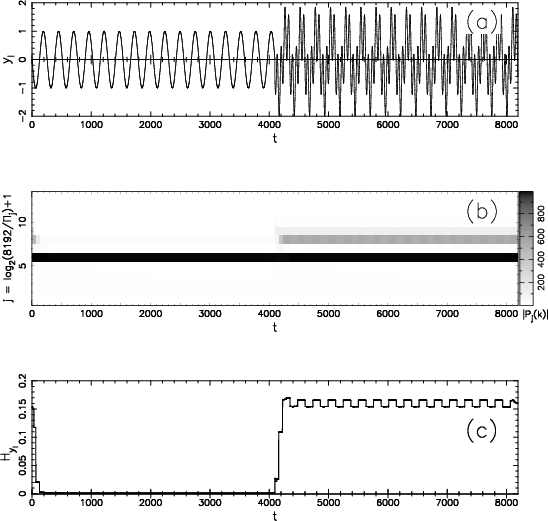}
\caption{Analysis of the time-dependent complexity of the toy model defined 
by Eq.\,\ref{Eqn:toy_model1}. From top to bottom: (a) time series, (b)
discrete wavelet transform, (c) discrete wavelet transform information
measure $H_{y_{I}}$.}
\label{Fig:toy_model1}
\end{figure}

%Caption Status: 2
%%%%%%%%%%%%%%%%%%%%%%%%%%%%%%%%%%%%%%%%%%%%%%%%%%%%%%%%%%%%%%%%%%%%%%%%%%%%%
%%%%%%%%%%%%%%%%%%%%%%%%%%%%%%%%%%%%%%%%%%%%%%%%%%%%%%%%%%%%%%%%%%%%%%%%%%%%%

% Figure 4: Toy Model2
%%%%%%%%%%%%%%%%%%%%%%%%%%%%%

\begin{figure}
\centering
\includegraphics[height=100mm,width=80mm]{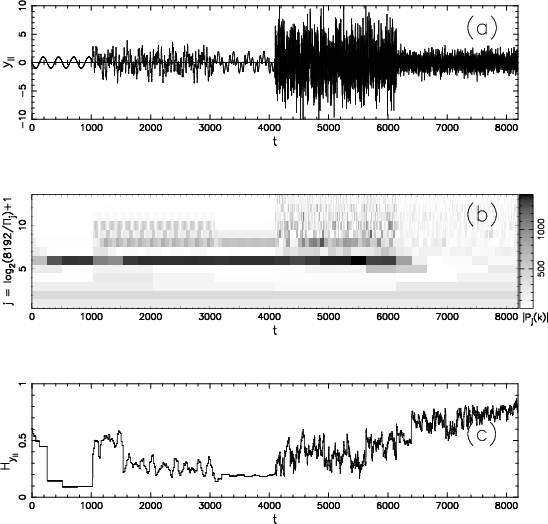}
\caption{Analysis of the time-dependent complexity of the toy model defined 
by Eq.\,\ref{Eqn:toy_model2}. From top to bottom: (a) time series, (b)
discrete wavelet transform, (c) discrete wavelet transform information
measure $H_{y_{II}}$.}
\label{Fig:toy_model2}
\end{figure}

%Caption Status: 2
%%%%%%%%%%%%%%%%%%%%%%%%%%%%%%%%%%%%%%%%%%%%%%%%%%%%%%%%%%%%%%%%%%%%%%%%%%%%%

We illustrate the capability of the method to capture time-dependent
changes in the dynamics with two toy models. Let us consider the two
time series $y_{I}(t_{q})$ and $y_{II}(t_{q})$
$(q=1,...,8192;\,t_{q}=q\Delta=q)$ constructed as follows:
\begin{equation}\label{Eqn:toy_model1}
y_{I}(t_{q})= \left\{ 
\begin{array}{lr}
y_{A}, & 0< q\leq 4096,\\
y_{B}, & 4096< q\leq 8192,
\end{array}
\right.
\end{equation}
and
\begin{equation}\label{Eqn:toy_model2}
y_{II}(t_{q})= \left\{ 
\begin{array}{lr}
y_{A}, & 0< q\leq 1024,\\
y_{C}, & 1024< q\leq 3072,\\
y_{B}, & 3072< q\leq 4096,\\
y_{D}, & 4096< q\leq 6144,\\
y_{E}, & 6144< q\leq 8192.\\
\end{array}
\right.
\end{equation}
The series $y_{I}$ and $y_{II}$ of Eqs.\,(\ref{Eqn:toy_model1}) and
(\ref{Eqn:toy_model2}) are made up of $2$ and $5$ different dynamical
regimes, respectively. Each is a different linear combination of
Fourier components and white noise:
\begin{eqnarray}
y_{A} & = & \sin\left( 2\pi\times \frac{32}{8192}q+\pi \right),\\
y_{B} & = & \sin\left( 2\pi\times \frac{32}{8192}q+\pi\right) \nonumber \\
& + & \sin\left( 2\pi\times \frac{128}{8192}q+\frac{5\pi}{4}\right),\\
y_{C} & = & \sin\left( 2\pi\times \frac{32}{8192}q
\right) + \sin\left( 2\pi\times \frac{128}{8192}q+\frac{\pi}{4}\right)\nonumber \\
 & + &\sin\left( 2\pi\times \frac{512/3}{8192}q+\frac{\pi}{8}\right)\nonumber \\
& + &\sin\left( 2\pi\times
\frac{8192/17}{8192}q+\frac{\pi}{17}\right), \\
y_{D} & = & y_{C} + \mathcal{G}_{3}(q),\\
y_{E} & = & \mathcal{G}_{1}(q).
\end{eqnarray}
Here $\mathcal{G}_{3}(q)$ and $\mathcal{G}_{1}(q)$ are Gaussian white
noise functions of amplitude unity, zero mean and an absolute variance
of $3$ and $1$, respectively. The complexity increases from one regime
to the next: as we go from regime $y_{A}$, through regime $y_{B}$ up
to $y_{C}$, we double the number of frequency components each time.
We then construct $y_{D}$ by adding the noise function
$\mathcal{G}_{3}(q)$ to the $y_{C}$ signal. Finally, regime $y_{E}$
consists of random numbers only and defines our signal of maximum
complexity, i.e. with no underlying periodic signal. In what follows
we study the performance of the DWaTIM to capture the time-dependent
complexity of the toy models $y_{I}$ and $y_{II}$.\\
\indent The complexity analysis for toy model $y_{I}$ of
Eq.\,(\ref{Eqn:toy_model1}) is represented in
Fig.\,\ref{Fig:toy_model1}. Figure \ref{Fig:toy_model1}(a) shows the
time series $y_{I}(t_{q})$. Until $t=4096$, the signal has the same
unique base frequency as the sinusoid $y_{1}$ of
\S\ref{Subsubsect:bfx}. However,
at $t=4096$ the behavior of the signal changes abruptly. A second
component, with a frequency $4$ times higher than the primary
component, is added to the signal. Figures
\,\ref{Fig:toy_model1}(b) and (c) show the associated complexity
analysis. The DWaT scalogram of Fig.\,\ref{Fig:toy_model1}(b)
illustrates the power spectrum of the base frequencies $1/\Pi_{j}$ as
a function of time. The variation of complexity with time of signal
$y_{I}$ is clearly identifiable, as the increase in complexity at
$t=4096$ is well recovered. The modes represented by scales $j=8$ and
$j=9$ show a response between $4096<t\leq 8192$. The related DWaTIM
complexity measure $H_{y_{I}}$ is plotted in
Fig.\,\ref{Fig:toy_model1}(c). Between $4096<t\lesssim 4200$ the
complexity rises from $H_{y_{I}}\simeq 6\cdot 10^{-4}$ to an average
value of $H_{y_{I}}\simeq 0.16$ for $t>4200$. We note that in the time
interval $4096<t\lesssim 4200$, $H_{y_{1}}$ rises progressively. The
size of this transient interval corresponds to about one half of the
period of $y_{A}$ which is $256$ units. This is the longest of the two
periods of the Fourier components involved in the transition
$y_{A}\rightarrow y_{B}$. This sets a limit on the time-localization
of the transition. In \S\ref{Subsect:pyth} we will discuss in detail the
latency of transitions in the dynamics measured by the DWaTIM in an
application to small-$N$ problems. We also remark that border effects
between $0\leq t\lesssim 200$ affect the measure for less than 3\% of
the signal.\\
\indent Figure \ref{Fig:toy_model2} shows the results for the second toy model
$y_{II}$ of Eq.\,(\ref{Eqn:toy_model2}). The related time series is
shown in Fig.\,\ref{Fig:toy_model2}(a) and the results for the
complexity analysis are represented in Fig.\,\ref{Fig:toy_model2}(b)
and (c). The time-dependent subtleties of the frequency spectrum are
depicted by the DWaT in Fig.\,\ref{Fig:toy_model2}(b). For instance,
the drop in complexity at $t\approx 3072$, where we remove the two
incommensurate base frequencies of regime $y_{C}$ and switch to regime
$y_{B}$, is recovered. The modes $j=10$ and $11$, excited between
$1024\lesssim t\lesssim 3072$ (the $j=11$ response during that
interval is hardly visible on Fig.\,\ref{Fig:toy_model2}[b]), do not
show a response any more between $3072\lesssim t\lesssim 4096$. The
increased complexity between $y_{C}$ and its noisy counterpart $y_{D}$
is also apparent; scales $7$ to $13$ show a more irregular pattern
between $4096\lesssim t
\lesssim 6144$ than between $1024\lesssim t\lesssim 3072$. 
\mybold{Finally, for the white-noise regime $y_{E}$ between $6144\lesssim t \lesssim
8192$, the DWaT response decreases. All the modes $j=1,...,13$ are on
average less excited than e.g., during regime $y_{D}$, and the DWaT
intensity is spread over these scales in an almost uniform
manner. This shows that the DWaT correctly identifies this regime as
noise.} We remark that border effects may explain the feature seen at
scale $j=6$ between $0\lesssim t\lesssim 1536$, namely, that the
magnitude of that mode increases and decreases until it stabilizes at
the end of that interval. Therefore border effects may here persist
for up to $1536/8192\simeq 18\%$ of the duration of the signal. This
is still less than the conservative estimate of $25\%$ we have given
in
\S\ref{Subsubsect:sinusoid_complexity}. Figure
\ref{Fig:toy_model2}(c) shows the DWaTIM $H_{y_{II}}$. Apart from the
interval $0<t\lesssim 1536$ where the border effects spoil the
result\footnote{The measure should give 0 between $0<t\lesssim
1024$. However, it is still true that the signal $y_{A}$ is quantified
as less complex than $y_{B}$ and that the transition from $y_{A}$ to
$y_{B}$ is picked up easily.}, the DWaTIM provides a consistent
overall diagnostic of time-dependent complexity. Each transition
between the regimes $y_{A}$ and $y_{E}$, through regimes $y_{C}$,
$y_{B}$ and $y_{D}$, is detected by the DWaTIM. For instance, the
transition at $t=3072$ from $y_{C}$ to $y_{B}$ corresponds to a drop
of about $0.05$ in average $H_{y_{II}}$ magnitude around that
instant. We recover a DWaTIM value for regime $y_{B}$ of
$H_{y_{II}}\approx 0.19$ between $3072\lesssim t\lesssim 4096$ i.e.,
of about $3\cdot 10^{-2}$ \mybold{units} larger than in toy model
$1$. This \mybold{19}\% difference is due to the response of the three
low-frequency scales $j=1$, $2$ and $3$ on Fig.\,\ref{Fig:toy_model2}
at these times; in toy model $1$ these modes do not show any response
(see Fig.\,\ref{Fig:toy_model1}[b] between $4096\lesssim t\leq
8192$). The performance of the DWaTIM to provide an absolute measure
of complexity for a given dynamical regime depends on the time series
subjected to analysis and on the number of scales included in the
DWaTIM computation. In \S\ref{Subsect:pyth} we investigate further
this effect for the case of $N$-body orbits. Finally, we illustrate
the behavior of the DWaTIM in the white-noise limit of $y_{E}$. As
mentioned earlier, the total wavelet energy between $6144 \lesssim
t\lesssim 8192$ (see grey-shade intensities on
Fig.\,\ref{Fig:toy_model2}[b]) is reduced with respect to the interval
$4096 \lesssim t\lesssim 6144$, pointing at the difficulty of the DWaT
to single out privileged frequencies with a high probability during
$6144 \lesssim t\lesssim 8192$. In this limit, the DWaTIM
asymptotically reaches its maximum value of $1$ (for the case of a
perfect white-noise signal and an infinite-length time series; see
\S\ref{Subsubsect:dwatim}).

\section{Few-Body Encounters}
\label{Sect:few-body}

% Remark: here we use bold typeface (unlike in method section) for the reviewer

%\newline
\noindent To validate the technique for the case of orbital dynamics,
 we present three applications to small $N$ problems ($N=2$, $3$ and
 $5$).  All individual orbits were computed using the \textbf{starlab}
 software environment \citep{starlab}. Integrations are performed
 using individual time-steps \citep{Aarseth85} and a fourth-order
 Hermite predictor-corrector scheme \citep{Makino92}. Time series are
 constructed by Hermite interpolation at evenly spaced time intervals
 and by projecting the orbit of each star on the three orthogonal axes
 $x$, $y$ and $z$ in the center of mass coordinate system. Standard
 $N$-body units are used throughout \citep{Heggie86}.

\subsection{Binary motion: $N$=2}
\label{Subsect:binaries}

\begin{figure*}
\vbox to170mm{
\vfil
\centering
\includegraphics[height=145mm,width=155mm]{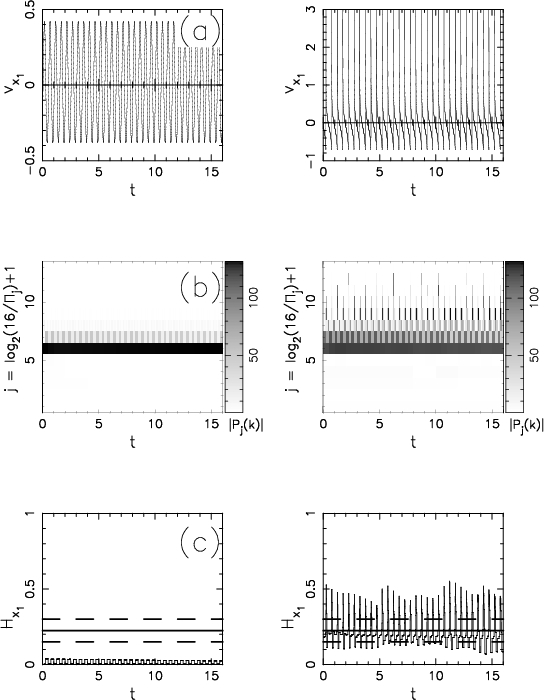}
\caption{Complexity measures of two unperturbed
  binaries with eccentricity $e=0.05$ (left-hand panels) and $e=0.95$
  (right-hand panels), respectively. For each binary, we show the
  $x$-component result of body \#1. From top to bottom: (a) velocity
  time series, (b) discrete wavelet transform (DWaT), (d) discrete
  wavelet transform information measure (DWaTIM) $H_{x_{1}}$. The solid
  horizontal lines in (c) show the average over
  the $x$-, $y$-, and $z$-components for $100$ binaries with $e$ and
  $\Pi_{j}$ randomly chosen between $0\leq e< 1$ and $2^{-8}\leq
\Pi_{j}\leq 2$, respectively. Dashed lines are one standard deviation.}
\label{Fig:binaries}
\vfil}
\end{figure*}

%\newline
\noindent We study independently the motion of two
unperturbed, equal-mass binaries with respective eccentricities
$e_{1}=0.05$ and $e_{2}=0.95$. Results for the $v_{x}$-component analysis
of body~\#1 of each binary are presented in the left-hand and
right-hand panels of Fig.\,\ref{Fig:binaries}, respectively. Figure
\ref{Fig:binaries}(a) shows the velocity time series $v_{x}$
sampled at $8192$ regular time intervals. The binaries are integrated
over $16$ $N$-body time units, so the sampling interval $\Delta$ is
$16/8192=2^{-9}$.  The period of both binaries is $0.5$, allowing a
sampling rate of $256$ data points per revolution. We note that a
consistent choice of $\Delta$ is of major importance to the method. If
an orbit contains frequency components that exceed $f_{\rm cr}$ (see
\S\ref{Subsubsect:definitions}) the DWaTIM will be aliased. A reasonable
selection of $\Delta$ must thus ensure that the complete dynamics of
the system is reproduced for the time-scales of interest.

The complexity diagnostics for the two binaries are given in
Fig.\,\ref{Fig:binaries}(b) and (c). The DWaTIM $H_{x_{1}}$ is
extracted from the DWaT (Fig.\,\ref{Fig:binaries}[b]) and is shown on
Fig.\,\ref{Fig:binaries}(c). The complexity measure oscillates in time
with the same frequency as the binary (in the same way than for the
$y_{I}$ toy model of Fig.\,\ref{Fig:toy_model1} for $t\gtrsim
4096$). The amplitude of these oscillations is greater for the more
eccentric binary. This is so because \mybold{the larger the
eccentricity of a binary the larger the variations in velocity}. In
the limit where the semi-major axis is sufficiently large, the
velocity of both bodies at apocenter is close to zero; likewise, the
complexity of any static configuration would asymptotically approach
zero. At apocenter we find, for the $e=0.05$ binary, a DWaTIM of
$\approx 0$ whereas, for the $e=0.95$ case, we obtain a value of
$\approx 0.1$. In contrast, at pericenter the velocities change
rapidly, requiring a broader frequency spectrum and thus implying a
higher complexity. Here the DWaTIM indicator increases to $\approx
0.03$ in the $e=0.05$ case and to $\approx 0.5$ for the $e=0.95$
binary. We note that a circular binary ($e=0$) has a constant velocity
modulus: for this case the complexity is zero throughout (for all the
regions not affected by border effects, see
\S\ref{Subsubsect:sinusoid_complexity}). We also remark that the time-averaged
DWaTIM would give approximately a constant value. This gives a way to
quantify an external perturbation acting on a stable Keplerian orbit
(see \S\ref{Subsect:pyth}).

We also computed the average complexity of the $v_{x}$-, $v_{y}$- and
$v_{z}$-DWaTIM measures for a total of $100$ different binaries. For each
binary, the eccentricity and the periodicity was randomly chosen in
the intervals $0\leq e< 1$ and $2^{-8}\leq
\Pi_{j}\leq 2$, respectively. The corresponding results for the DWaTIM
 are shown by the horizontal lines (in blue) on
 Fig.\,\ref{Fig:binaries}(c). The solid central line displays the mean
 DWaTIM; the interval delimited by the upper and the lower dashed line
 indicates the standard deviation. The average DWaTIM is $\approx
 0.22$. This value can be seen as indicative for the complexity of
 unperturbed binaries with a periodicity $\Pi_{j}$ comprised within
 the DWaT bandwidth.

%%%%%%%%%%%%%%%%%%%%%%%%%%%%%%%%%%%%%%%%%%%%%%%%%%%%%%%%%%%%%%%%%%%%%%%%%%%%%

% Figure 5: Binaries
%%%%%%%%%%%%%%%%%%%%%%%%%%%%%

%Caption Status: 2
%%%%%%%%%%%%%%%%%%%%%%%%%%%%%%%%%%%%%%%%%%%%%%%%%%%%%%%%%%%%%%%%%%%%%%%%%%%%%

\subsection{The Pythagorean problem: $N$=3}
\label{Subsect:pyth}

\begin{figure}
\includegraphics[height=50mm,width=80mm]{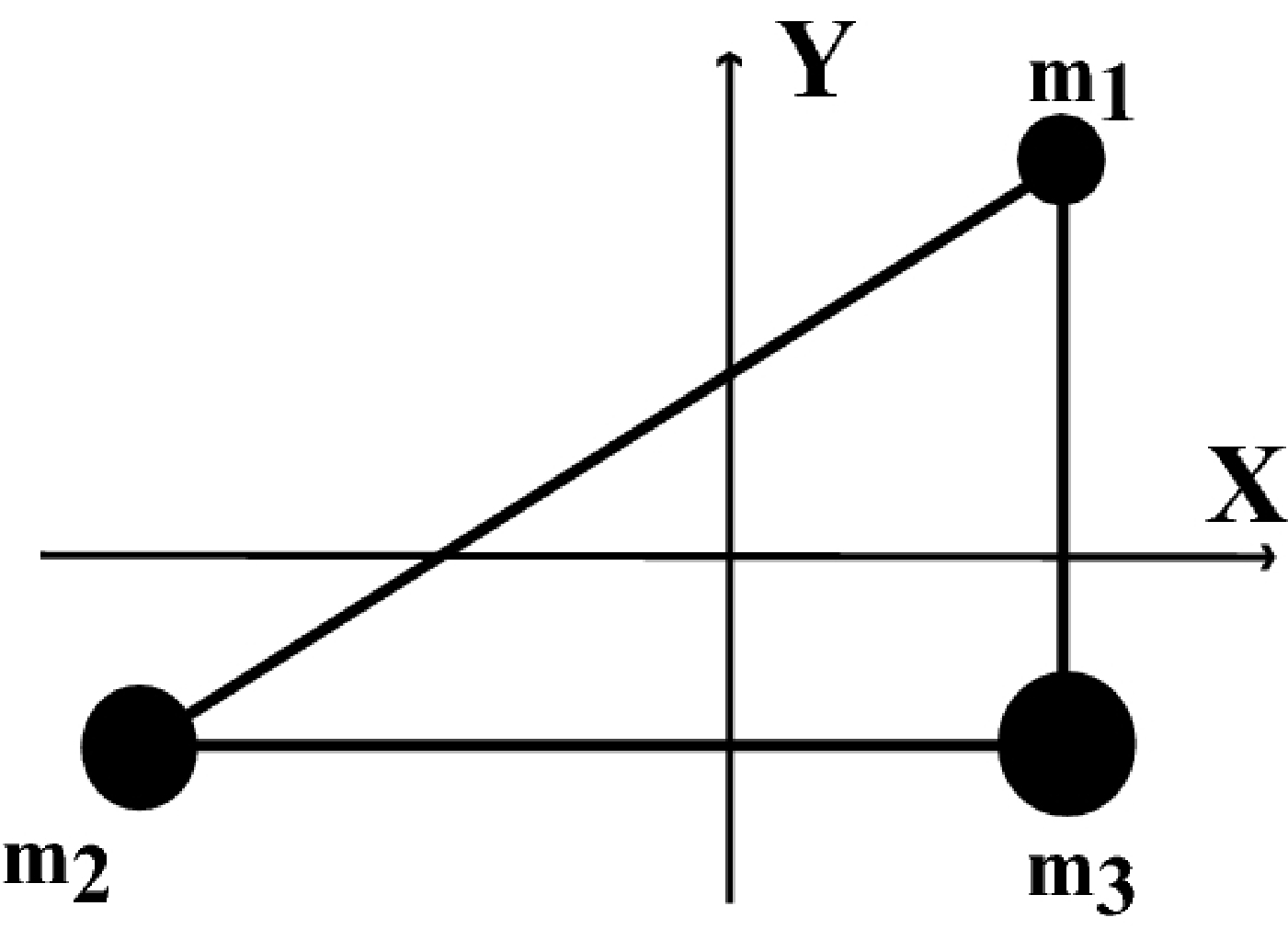}
\caption{Pythagorean problem: initial configuration.}
\label{Fig:pyth_scheme}
\end{figure}

\begin{figure*}
\vbox to170mm{
\vfil
\centering
\includegraphics[height=145mm,width=155mm]{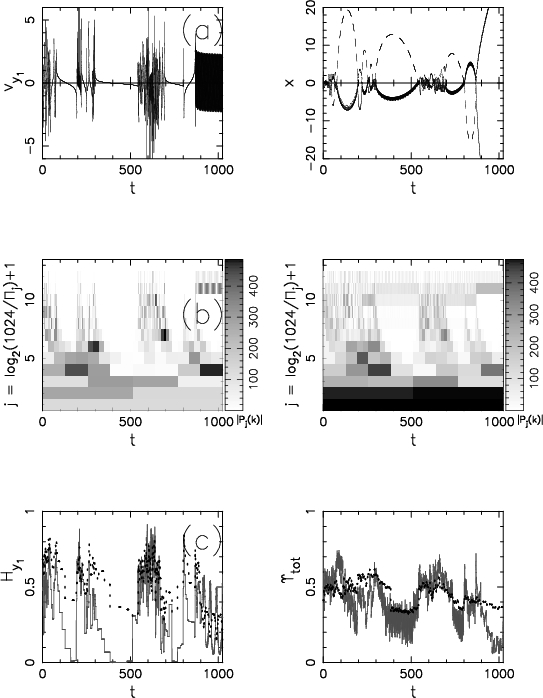}
\caption{Complexity measures of the $x$ component of particle \#1
  (left-hand panels) and averaged overall complexity (right-hand
  panels) of the Pythagorean problem (see Table \ref{Tab:Pyth},
  Fig.\,\ref{Fig:pyth_scheme} and text for further detail). From top
  to bottom: (a) velocity time series (left-hand panel) and $x$
  position time series of the $3$ particles (right-hand panel), (b)
  discrete wavelet transform (DWaT) (left-hand panel) and average DWaT
  (right-hand panel), (c) discrete wavelet transform information
  measure (DWaTIM) $H_{y_{1}}$ and average DWaTIM
  $\Upsilon_{\rm tot}$. The \mybold{dotted} lines on (c) are obtained by
  computing the measures using the complete bandwidth of the DWaT. The
  solid lines show the same result computed by excluding the
  contribution of scales $1,...,4$.}
\label{Fig:pyth}
\vfil}
\end{figure*}

%\newline
\noindent The well-studied Pythagorean configuration
\citep{Burrau13} is a classic example of long-term complex behavior
(see e.g., \citealt{Aarseth94}). The initial conditions consist of
three particles at rest, placed at the vertices of a Pythagorean
triangle. The initial conditions are given in Table \ref{Tab:Pyth} and
the configuration is depicted in Fig.\,\ref {Fig:pyth_scheme}.\\
\indent Orbital integrations were performed over $1500$ time units by
repeatedly re-running the initial configuration with a reduced
per-step integration error until convergence of the result was
reached. In this way, the final trajectories showed a total absolute
energy error $|E_{\rm final}-E_{\rm initial}|$ of less than
$10^{-10}$. The analysis is restrained to the first $1024$ time units
of integration, during which we found by visual inspection that the
system spent a comparable amount of time in trivial and in more
complicated states. Once more, the time series comprise $Q=8192$ data
points. Results of the respective time series analysis for particle
\#1 and for the entire Pythagorean problem are shown in the left-hand
and right-hand panels of Fig.\,\ref{Fig:pyth}, respectively.\\
%%%%%%%%%%%%%%%%%%%%%%%%%%%%%%%%%%%%%%%%%%%%%%%%%%%%%%%%%%%%%%%%%%%%%%%%%%%%%
%
% Table 1: Pythagorean initial conditions
%%%%%%%%%%%%%%%%%%%%%%%%%%%%%
\begin{table}%[htbp]
\begin{tabular}{cccccccc}%{||l||p{1cm}|p{0.25cm}|p{0.25cm}|p{0.25cm}|p{0.25cm}|p{0.25cm}|p{0.25cm}||}
\hline
Body & mass & $x$ & $y$ & $z$ & $v_{x}$ & $v_{y}$ & $v_{z}$\\
\hline
\#1 & 3 & 1 & 3 & 0 & 0 & 0 & 0\\
\#2 & 4 & $-2$ & $-1$ & 0 & 0 & 0 & 0\\
\#3 & 5 & 1 & $-1$ & 0 & 0 & 0 & 0\\
\hline
\end{tabular}
\caption{Pythagorean problem: initial conditions.}
\label{Tab:Pyth}
\end{table}
%Caption Status: 2
%%%%%%%%%%%%%%%%%%%%%%%%%%%%%%%%%%%%%%%%%%%%%%%%%%%%%%%%%%%%%%%%%%%%%%%%%%%%%
%%%%%%%%%%%%%%%%%%%%%%%%%%%%%%%%%%%%%%%%%%%%%%%%%%%%%%%%%%%%%%%%%%%%%%%%%%%%%
%
% Figure 6: Pythagorean scheme
%%%%%%%%%%%%%%%%%%%%%%%%%%%%%
%
%Caption Status: 2
%%%%%%%%%%%%%%%%%%%%%%%%%%%%%%%%%%%%%%%%%%%%%%%%%%%%%%%%%%%%%%%%%%%%%%%%%%%%%
%
% Figure 7: Pythagorean problem
%%%%%%%%%%%%%%%%%%%%%%%%%%%%%
%
%
%Caption Status: 2
%%%%%%%%%%%%%%%%%%%%%%%%%%%%%%%%%%%%%%%%%%%%%%%%%%%%%%%%%%%%%%%%%%%%%%%%%%%%%
%
\indent The $v_{y}$-coordinate of body \#1 is shown in the left-hand
Fig.\,\ref{Fig:pyth}(a). In the right-hand panel of
Fig.\,\ref{Fig:pyth}(a) we illustrate the dynamics of the whole
Pythagorean problem by showing the position $x$ of all the $3$
bodies.  Here we see that the system is characterized by a highly
complicated interplay of the three particles, including a sequence of
intermittent binary formation and disruption. In the time interval
$300 \lesssim t
\lesssim 540$ for instance, the single body \#1 strides away to large
distance on a smooth trajectory (upper \mybold{dashed} line in the right-hand
Fig.\,\ref{Fig:pyth}[a] during that interval). Due to the recoil, a
binary, formed of bodies
\#2 and \#3, leaves in the opposite direction. The motion of the
binary stars about their common centre of mass gives the thick line
seen in the lower part of the right-hand Fig.\,\ref{Fig:pyth}(a)
between $300\lesssim t \lesssim 540$.\\
\indent The DWaT scalogram of body \#1 and the average DWaT
of the Pythagorean problem (the average magnitude of base frequency
$1/\Pi_{j}$ at integration time $t$), computed by averaging over the
$3$ particles and over the $x$- and $y$-components, are shown in the
left-hand and right-hand panels of Fig.\,\ref{Fig:pyth}(b),
respectively. The DWaTIM for body \#1, $H_{y_{1}}$, and the overall
DWaTIM of the Pythagorean problem, $\Upsilon_{\rm tot}$ (see
\S\ref{Subsubsect:dwatim}), are shown in
Fig.\,\ref{Fig:pyth}(c). The \mybold{dotted} line depicts the DWaTIM as
computed by taking into account all the scales $j=1,...,13$. The solid
line highlights the dramatic
improvement of the diagnostic, obtained when ignoring the first $4$
scales ($j=1,...,4$) in the computation. As mentioned earlier (see
\S\ref{Subsubsect:sinusoid_complexity}), both complexity measures cease 
to yield reliable results in the limit where the orbit is
resolved over less than $8$ complete oscillations (i.e., dynamical
times). This motivated our decision to exclude the low-frequency DWaT
scales in a general manner. We take the conservative choice of not
including the $j=4$ scale (i.e. the periodicities $\Pi_{4}$
corresponding to $1/8$ of the full signal). For the
remainder of this work, we only compute and discuss this improved
version of the DWaTIM.
\newline\indent The performance of the DWaTIM can be studied with respect to two
criteria: 1) time-localization, i.e. the ability of the measure to
identify accurately the instants $t$ where qualitative changes in the
orbital dynamics occur, and 2) continuous quantification of
complexity, i.e. the ability of the measure to provide a consistent
evaluation of complexity in an uninterrupted manner. In order to
discuss 1), we examine the time series of body \#1 of the left-hand
Fig.\,\ref{Fig:pyth}(a). Let us consider the transitions from a regime
where body \#1 approaches bodies \#2 and \#3 on a comparatively smooth
trajectory to the more complex regime where it undergoes multiple
close encounters with the latter two bodies. By visual inspection of
the left-hand and right-hand panels on Fig.\,\ref{Fig:pyth}(a), we
identify such transitions to occur at e.g., $t\approx 200$, $t\approx
530$ and $t\approx 800$. In each of these cases, \mybold{we can see in
Fig.\,\ref{Fig:pyth}(c) that the measure $H_{y_{1}}$ shows the
corresponding transition.} For example, the transition at $t\approx
530$ is identified by the DWaTIM by an increase of about $0.7$. The
instantaneous close encounter between the $3$ particles at $t\approx
800$ is also singled out. Here we find a local peak of
$H_{y_{1}}\approx 0.8$. Likewise, the \mybold{instant} $t\approx 875$
\mybold{at which} body
\#1 forms a binary system with one of the remaining bodies 
is \mybold{also} recovered; the complexity shows an instantaneous peak
at that instant and then gradually decreases from $H_{y_{1}}\approx
0.6$ at $t\approx 875$ to a mean value of $H_{y_{1}}\approx 0.25$ at
$t=1024$. The accuracy with which the epoch of a transition is
resolved depends on the frequency bands involved in the transition. As
mentioned earlier, the DWaT resolution in time increases by a factor
of $2$ when shifting from scale $j$ to $j+1$. High-frequency
transitions can therefore be localized more accurately in time than
low-frequency transitions. The maximum latency of the DWaTIM is
obtained by studying the particular case of a transition involving the
two lower scales $j=5$ and $6$. The maximum latency then corresponds
to the resolution at scale $j=6$, i.e. $1024/2^{5}=32$ time
units. This sets an upper limit to the error on the DWaTIM performance
in time-localization. In conclusion, qualitative changes of orbital
dynamics can be accurately localized in time by the DWaTIM
indicator.\\
\indent Concerning 2) the continuous quantification of complexity, we argue
that, broadly speaking, the $H_{y_{1}}$ of the left-hand
Fig.\,\ref{Fig:pyth}(c) reproduces the complexity of the frequency
spectrum obtained by the DWaT in Fig.\,\ref{Fig:pyth}(b) in a
consistent manner. The relative complexity of the time series is
quantified continuously in time with respect to the two limiting cases
of a single base-frequency sinusoid with $H_{y_{1}}=0$ and a white
noise signal of $H_{y_{1}}=1$. Excluding the low-frequency scales
$j=1,...,4$ has also some bearings on the reliability of the
diagnostics. For instance, when these frequencies are included, the
DWaTIM produces inconsistent results in the interval $300\lesssim
t\lesssim 540$ (see dash-dotted line in Fig.\,\ref{Fig:pyth}[c]).  The
complexity in the case of unperturbed motion of body \#1 is then
estimated to be higher than for the case where the body is bound in a
binary (compare to the dash-dotted line in the interval $875\lesssim t
\lesssim 1024$).\\
\indent We also investigated the effects of analyzing the motion of body \#1
through a different time series on the $H_{y_{1}}$ complexity
diagnostic. This additional consistency check allows to measure DWaT
border effects for the case of the Pythagorean problem and to examine
the extent to which the DWaTIM is able to provide an absolute measure
of complexity. To compare two time series we constructed a new one by
taking the first half of the $Q=8192$ points signal of
Fig.\,\ref{Fig:pyth}(a), i.e. up to $t=512$. This gave us a new time
series $v'_{y_{1}}(t_{q})$ of $Q'=4096$ data points. The DWaT analysis
was then performed over a different frequency domain than for the
original $Q=8192\,(t=1024)$ case.  The time series $v'_{y_{1}}(t_{q})$
had also a different end value than its parent time series
$v_{y_{1}}(t_{q})$, i.e. $v'_{y_{1}}(t_{4096})\neq
v_{y_{1}}(t_{8192})$. The amplitude of DWaT border effects were
therefore different than for the $Q=8192$ case (see
\S\ref{Subsubsect:bfx}). We computed the DWaTIM $H'_{y_{1}}$ and
compared it with $H_{y_{1}}$ shown on Fig.\,\ref{Fig:pyth}(c) in the
interval $0\leq t\leq 512$. Except for values right at the edge of
Fig.\,\ref{Fig:pyth}(c), the results were nearly identical. In
particular, the rms residual between the two measures was $\approx
0.095$ between $0\leq t\leq 128$ (corresponding to the first 25\% of
the signal $v'_{y_{1}}$) and $\leq 10^{-3}$ between $128<t\leq 512$.
\newline\indent  We obtain similar results for the time series of
the other two bodies \#2 and \#3, and for the $x$-component. For that
reason we omitted to display those results. The overall DWaTIM
$\Upsilon_{\rm tot}$ of the Pythagorean problem (shown on
Fig.\,\ref{Fig:pyth}[c], right-hand panel) gives a time-resolved
insight on the global orbital evolution obtained by integrating the
initial conditions of Table
\ref{Tab:Pyth}. Once more, the features seen in the curve of
$\Upsilon_{\rm tot}$ match those in real space of Fig.\,\ref{Fig:pyth}(a),
right-hand side panel. We further discuss the utility of this general
measure in the next section \S\ref{Subsect:caledonian}.

%%%%%%%%%%%%%%%%%%%%%%%%%%%%%%%%%%%%%%%%%%%%%%%%%%%%%%%%%%%%%%%%%%%%%%%%%%%%%

% Figure 8: Caledonian 5-body problem: initial configuration
%%%%%%%%%%%%%%%%%%%%%%%%%%%%%
\begin{figure}
%\centering
\includegraphics[height=60mm,width=80mm]{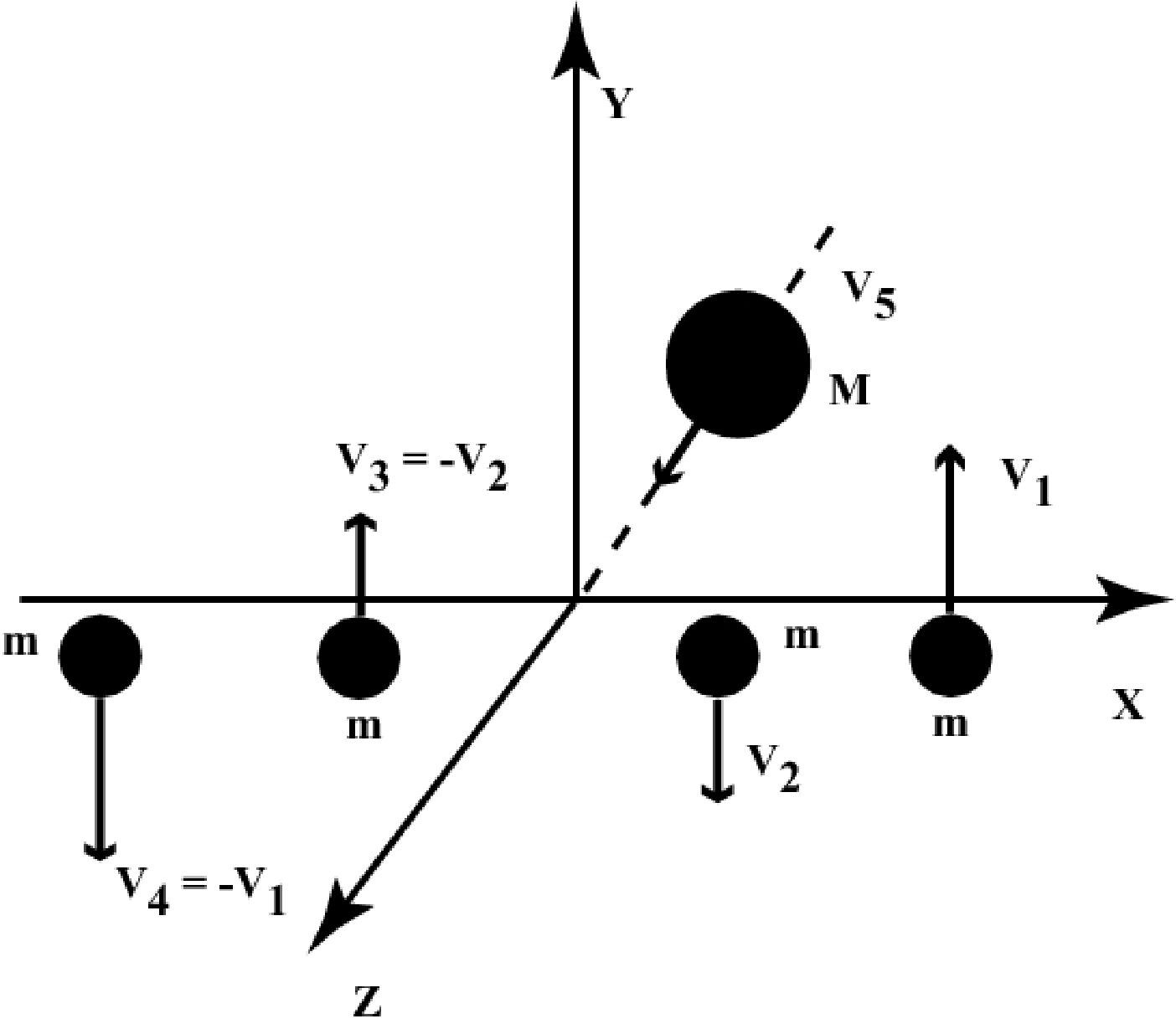}
\caption{Caledonian 5-body problem: initial configuration.}
\label{Fig:caledonian_scheme}
\end{figure}

%Caption Status: 2
%%%%%%%%%%%%%%%%%%%%%%%%%%%%%%%%%%%%%%%%%%%%%%%%%%%%%%%%%%%%%%%%%%%%%%%%%%%%%

\subsection{Perturbed $N$=5 Caledonian configuration}
\label{Subsect:caledonian}

Next we perturb a configuration of the planar Caledonian symmetric
four-body problem (hereafter CP;
\citealt{Szell04}). Initial conditions are shown in
Fig.\,\ref{Fig:caledonian_scheme} and in Table \ref{Tab:Cal}. The CP
is set up as described in Figure $1$ of
\citet{Szell04}. We specialize to the case where all bodies have masses equal to $1$. For
this configuration, the unperturbed CP consists of two stable binaries
evolving in the $x-y$ plane around their common center of mass. We now
perturb the CP configuration with a fifth body of mass $=1.5$ (body
\#5 in Table\,\ref{Tab:Cal}). The perturber approaches the $x-y$ plane
as it moves towards the center of mass of the two binaries on a time
scale of $\approx 8\pi/5$. \mybold{At that moment} the distance
between body
\#5 and the center of mass is $\approx 0.15$ i.e., about the
separation of the two binaries. Therefore the CP is strongly perturbed
from $t\approx 5$ onwards. The subsequent evolution is characterized
by repeated interactions including the formation of hierarchical
systems with several particles orbiting around a hard central binary.

%%%%%%%%%%%%%%%%%%%%%%%%%%%%%%%%%%%%%%%%%%%%%%%%%%%%%%%%%%%%%%%%%%%%%%%%%%%%%

% Table 2: Caledonian problem: initial conditions
%%%%%%%%%%%%%%%%%%%%%%%%%%%%%
\begin{table}%[htbp]
\begin{tabular}{cccccccc}%{||p{0.3cm}||p{0.3cm}|p{1.0cm}|p{0.2cm}|p{1.0cm}|p{0.2cm}|p{1.0cm}|p{0.7cm}||}
\hline
Body & mass & $x$ & $y$ & $z$ & $v_{x}$ & $v_{y}$ & $v_{z}$\\
\hline
\#1 & 1 & 0.55 & 0 & 0.27 & 0 & 0.23 & 0.00\\
\#2 & 1 & 0.30 & 0 & 0.27 & 0 & $-2.42$ & 0.00\\
\#3 & 1 & $-0.30$ & 0 & 0.27 & 0 & 2.42 & 0.00\\
\#4 & 1 & $-0.55$ & 0 & 0.27 & 0 & $-0.23$ & 0.00\\
\#5 & 1.5 & 0.00 & 0 & $-0.72$ & 0 & 0.00 & 0.03\\
\hline
\end{tabular}
\caption{Caledonian 5-body problem: initial conditions.}
\label{Tab:Cal}
\end{table}
%Caption Status: 2
%%%%%%%%%%%%%%%%%%%%%%%%%%%%%%%%%%%%%%%%%%%%%%%%%%%%%%%%%%%%%%%%%%%%%%%%%%%%%

%%%%%%%%%%%%%%%%%%%%%%%%%%%%%%%%%%%%%%%%%%%%%%%%%%%%%%%%%%%%%%%%%%%%%%%%%%%%%

% Figure 9: Caledonian problem (large, DEFUNCT)
%%%%%%%%%%%%%%%%%%%%%%%%%%%%%
%\begin{figure*}
%\vbox to170mm{
%\vfil
%\centering
%\includegraphics[height=100mm,width=80mm]{revised_images/caledonian_nbody_sh1p_dwatim.jpg}
%\caption{Overall complexity of
%the Caledonian problem (see also Table \ref{Tab:Cal} and
%Fig.\,\ref{Fig:caledonian_scheme}). From top to bottom: (a) position
%time-series of the $5$ particles, (b) averaged discrete wavelet
%transform (DWaT), (c) overall discrete wavelet transform information
%measure (DWaTIM) $\Upsilon_{\rm tot}$.}
%\label{Fig:caledonian}
%\vfil}
%\end{figure*}

%Caption Status: 2
%%%%%%%%%%%%%%%%%%%%%%%%%%%%%%%%%%%%%%%%%%%%%%%%%%%%%%%%%%%%%%%%%%%%%%%%%%%%%
%%%%%%%%%%%%%%%%%%%%%%%%%%%%%%%%%%%%%%%%%%%%%%%%%%%%%%%%%%%%%%%%%%%%%%%%%%%%%

% Figure 9: Caledonian problem
%%%%%%%%%%%%%%%%%%%%%%%%%%%%%

\begin{figure}
\centering
\includegraphics[height=100mm,width=80mm]{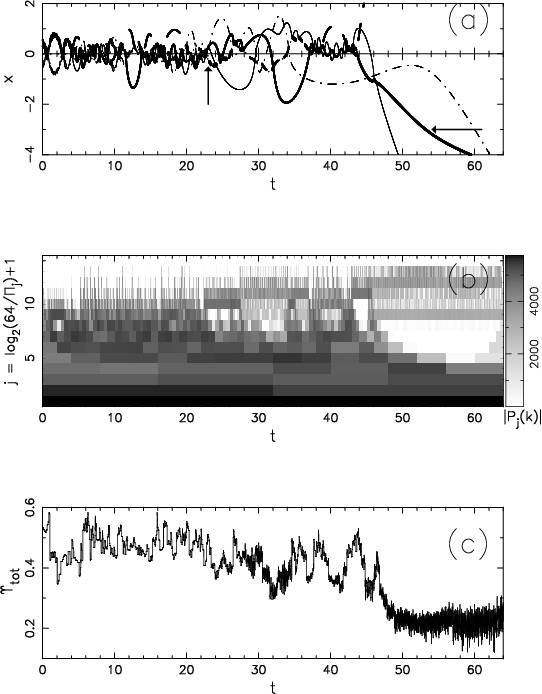}
\caption{Overall complexity of
the Caledonian problem (see also Table \ref{Tab:Cal} and
Fig.\,\ref{Fig:caledonian_scheme}). From top to bottom: (a) position
time-series of the $5$ particles, (b) averaged discrete wavelet
transform (DWaT), (c) overall information measure $\Upsilon_{\rm tot}$.}
\label{Fig:caledonian}
\end{figure}

%Caption Status: 2
%%%%%%%%%%%%%%%%%%%%%%%%%%%%%%%%%%%%%%%%%%%%%%%%%%%%%%%%%%%%%%%%%%%%%%%%%%%%%

\indent The outcome of the complexity analysis is summarized in
Fig.\,\ref{Fig:caledonian}. Figure \ref{Fig:caledonian}(a) gives the
$x$-position time series of the five particles; panels (b) and (c)
show the average DWaT and $\Upsilon_{\rm tot}$ (similarly to the
right-hand Figs.\,\ref{Fig:pyth}[a], [b] and [c]). Figure
\ref{Fig:caledonian}(a) gives an overview of the motion for the entire integration
time of $64$ time units. \mybold{(Note that the motion of particle
\#5, plotted with a dotted line on Fig.\,\ref{Fig:caledonian}[a],
 is difficult to disentangle and is hardly visible.)}  The intricate
 gravitational interplay between the five particles is difficult to
 follow by visual inspection: the individual evolution of the $5$
 orbits will not be discussed in detail. We focus on the evolution of
 the system as a whole. The dynamics up to $t\approx43$ may be
 described by roughly two qualitatively different states of
 motion. The first one is sometimes referred to as ``hierarchical
 interplay'' (see e.g., Gemmeke et al. 2006). It is characterized by a
 central binary and three particles orbiting at large radii. A clear
 example of this can be seen on Fig.\,\ref{Fig:caledonian} in the time
 interval $23\lesssim t\lesssim 36$. (The hard central binary forms at
 $t\approx 23$ as indicated by the vertical arrow). The second state
 of motion is what we call ``democratic interplay''. This regime is
 characterized by multiple close encounters between the $5$ bodies,
 each body contributing an approximately equal amount to the overall
 complexity of the system. Examples of democratic interplay are the
 time intervals $5\lesssim t\lesssim 10$ and $14\lesssim t\lesssim
 23$. At $t\approx 43$, the dynamics of the perturbed CP changes
 dramatically. After a close encounter around that time, all the
 particles stride away to larger distance. One particle immediately
 escapes the system on a nearly rectilinear trajectory (see upper
 dashed line on Fig.\,\ref{Fig:caledonian}[a] at that instant). The
 remaining bodies stay close to each other until $t\approx 46$, when
 they are subjected to a further close encounter and another body is
 ejected (see lower solid line). The subsequent motion is relatively
 smooth. One particle remains (dashed line) and orbits around a hard
 binary (indicated by the horizontal arrow).\\
\indent The overall complexity of the perturbed CP is shown by the DWaTIM in
Fig.\,\ref{Fig:caledonian}(c). Let us consider two time intervals,
prior to and after $t=43$.  When $t\leq 43$ the DWaTIM
$\Upsilon_{\rm tot}$ is roughly constant and it is interesting to observe
that the local minima seen in that time interval can be found during
the more quiescent hierarchical regimes, such as for example at
$t\approx 32$. The two peaks observed at $t\approx 43$ and $\approx
46$ arise from the two independent high-energy encounters that we have
described earlier taking place at those times. When the final binary
forms and the system dissolves the motion becomes unmistakably less
complex and consequently $\Upsilon_{\rm tot}$ has a lower value of
$\approx 0.2$ on average for all time $t\gtrsim 48$.

\section{Equal mass N=256 Plummer sphere} 
\label{Sect:plummer}

We apply the DWaTIM technique to a self-gravitating spherical
polytrope of index $n = 5$ (\citealt{Plummer11}; \citealt{BT87},
\S4.4.3).  The phase-space distribution function $\mathcal{F}$ of this
system is a power-law of $\varepsilon$,
\begin{equation}\label{Eqn:Plummer_distr}
\mathcal{F}(\varepsilon) =  \frac{24\sqrt{2}}{7\pi^3} \frac{R^2}{G^5M^4}(- \varepsilon)^{\frac{7}{2}} 
\end{equation}
where $\varepsilon = \frac{1}{2} v^2 + \Phi(r) < 0$ is the mechanical
energy per unit mass, $v$ the three-dimensional velocity, $\Phi$ the
potential which is a function of the radius $r$ only. Note that
$\mathcal{F} = 0$ by construction whenever $\varepsilon > 0$ so that
only mass elements bound by gravity are considered. \mybold{Given a
value of the} gravitational constant $G$, the two free parameters $M$
and $R$ define the total system mass and a reference unit of length,
respectively. Integrating $\mathcal{F}$ over all velocities at
constant radius yields the mass density $\varrho$ at that radius,
\begin{equation}\label{Eqn:Plummer_density}
\varrho(r) = \int_0^{\sqrt{-2\Phi(r)}} 4\pi\mathcal{F}(\varepsilon)  v^2 {\rm d}v 
= \frac{3}{4\pi}
\frac{M}{R^3}\left(1+\left[\frac{r}{R}\right]\hh{2}\right)\hh{-5/2}
\end{equation}
\mybold{where we have used} ${\rm d}\hh{3} \bmath{v} = 4\pi v^2{\rm d}v$ \mybold{valid for an
 isotropic velocity field and where we} have substituted for $\Phi(r)$
 by solving Poisson's equation.  Note that the length $R$ defines the
 radius of a \mybold{uniform-density core ($\varrho[r<<R] \approx $
 constant)}.  An $N = 256$ particle representation of
 Eq. (\ref{Eqn:Plummer_density}) is obtained by random-sampling the
 mass density to assign three-dimensional positions. All bodies have
 mass $m_i = M/N$, where $ i = 1,...,N$.  The particles' energy is
 attributed similarly from Eq.\,(\ref{Eqn:Plummer_distr}) from which
 we compute the square velocity $v^2 = 2( \varepsilon - \Phi[r])$ as
 in the standard method of
\citet{Aarseth74}.  In those circumstances the \mybold{total} kinetic and
gravitational energies satisfy the virial theorem of equilibrium systems.\\
\indent The nominal dynamical time is $t_{\rm d} = 2R/\sigma$ 
\mybold{where $\sigma^2$ is the mean squared velocity} averaged 
over mass \mybold{up to the half-mass radius. We found
$\sigma\hh{2}=0.392...GM/R \simeq \pi GM / 8 R$.}  The two-body
relaxation time is conveniently defined as $t_{\rm r} = N t_{\rm d} /
\ln 0.4 N $ (see
\citealt{BT87}). If we set Heggie-Mathieu
computational units with $G = M = 1$ and $R = 1.1781$ such that the
total binding energy $E = -1/4$, we \mybold{obtain $t_{\rm d} \approx
3.46$} so that for $N = 256$ bodies the relaxation \mybold{time
$t_{\rm r} \approx 191$} time units. For low-$N$ systems such as this
one, the diffusion of kinetic energy leads to core-collapse on roughly
that time scale, when $\varrho$ becomes singular at the center. It is
around that time that hard binaries form and energy exchanges between
bodies is at its most extreme. We therefore evolved the system for a
total time of 256 time units to ensure that the cluster reaches
core-collapse and contrast pre- and post-collapse evolution. The
equations of motion were integrated with no softening of the
potential.

%%%%%%%%%%%%%%%%%%%%%%%%%%%%%%%%%%%%%%%%%%%%%%%%%%%%%%%%%%%%%%%%%%%%%%%%%%%%%

% Figure 10: Lagrangian radii
%%%%%%%%%%%%%%%%%%%%%%%%%%%%%
\begin{figure}
\centering
\includegraphics[height=90mm,width=85mm]{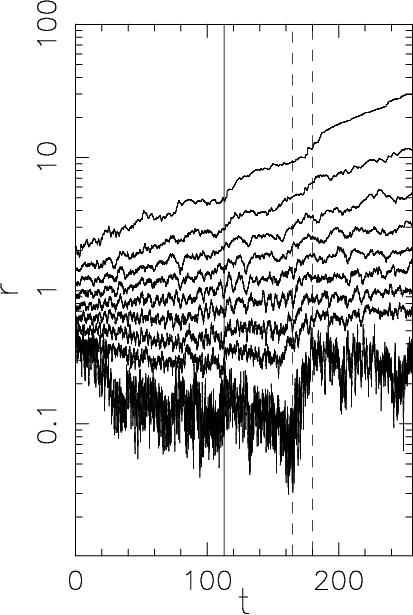}
\caption{From bottom to top: core radius $r_{\rm c}$, 20\%, 30\%,
  40\%, 50\%, 60\%, 70\%, 80\% and 90\% Lagrangian radii of the
  $N=256$ equal-mass Plummer sphere. The solid vertical line at
  $t=113$ marks the formation of the first hard binary in the
  system. The dashed vertical lines indicate the onset and the end of
  a notable core expansion between $165\lesssim t \lesssim 180$.}
\label{Fig:Plummer_lagrad}
\end{figure}

%Caption Status: 1
%%%%%%%%%%%%%%%%%%%%%%%%%%%%%%%%%%%%%%%%%%%%%%%%%%%%%%%%%%%%%%%%%%%%%%%%%%%%%

The velocities of all bodies are stored at regular time intervals and
the DWaTIM is computed in a post-simulation analysis.  The sampling
interval, $\Delta$, should be sufficiently small to capture the
dynamics in the dense core fully. A naive strategy would be to store
all data at all integration time steps. For the configuration adopted
here we have observed that the time steps may become as small as
$2^{-32} \simeq 2.32 \cdot 10^{-10}$; unfortunately a sampling rate of
that order would translate to a prohibitively high volume of data
storage even for modest values of $N$. Instead, we note that the
dynamical time $t_{\rm d}$ in principle sets a standard from which to pick
an adequate value of $\Delta$.  But because the central region becomes
ever denser during evolution, $t_{\rm d} \approx 3.46$ set from the initial
configuration would give no useful reference.

Instead, we explore the evolution in time of the mass profile of the
system and look for a minimum core length and velocity dispersion
inside that core.  The density-averaged core radius $r_{\rm c}$ defines a
quantity which monitors the rise of the central density (see e.g. von
Hoerner 1960, Casertano \& Hut 1985). From
Eq.\,(\ref{Eqn:Plummer_density}) we find for the initial configuration
$r_{\rm c} =R/2 \simeq 0.59$ (enclosing 13\% of the mass, or 33
bodies). Figure \ref{Fig:Plummer_lagrad} displays the time-evolution
of $r_{\rm c}$ and several Lagrange radii.  The core radius $r_{\rm c}$ always
appears at the bottom on Fig.\,\ref{Fig:Plummer_lagrad}. Up to $t
\approx 113$ units, $r_{\rm c}$ decreases on the mean. At that time, the
first hard binary of binding energy $E < -100 kT$ forms\footnote{The
energy scale $kT$ is defined by the condition that the total stellar
kinetic energy of the system, excluding internal binary motion, is
$3/2 NkT$.}.  That event is marked with a vertical full line on
Fig.\,\ref{Fig:Plummer_lagrad}. Note that $r_{\rm c}$ along with the 20\%
and 30\% Lagrange radii increase significantly from $t \approx
165$ onwards. This phase of rapid expansion indicates the on-set of
post-collapse evolution for that simulation. The minimum value of $r_{\rm c}
\simeq 0.05$ occurs at $t
\simeq 165$ units and encloses 3\% of the total mass (8 bodies).  We
compute a dynamical time $t_{\rm d,c} \simeq 0.05$ for the core at
that time, \mybold{a factor $\simeq 70$} smaller that the value
computed from the initial conditions. An orbit confined to the core is
adequately sampled with five points or more and hence we set $\Delta =
0.01$ for the complexity analysis. The time-resolution of any features
seen in the diagnostic of complexity is therefore $2\Delta = 0.02$
units, and any binary formed through dynamical evolution is well
sampled provided its binding energy $E > -2.75 kT$.

%%%%%%%%%%%%%%%%%%%%%%%%%%%%%%%%%%%%%%%%%%%%%%%%%%%%%%%%%%%%%%%%%%%%%%%%%%%%%

% Figure 11: Particle 6 and 127
%%%%%%%%%%%%%%%%%%%%%%%%%%%%%%%
\begin{figure*}
\vbox to140mm{ \vfil \centering
\centering
\includegraphics[angle=0,angle=0,angle=0,height=60mm,width=160mm]{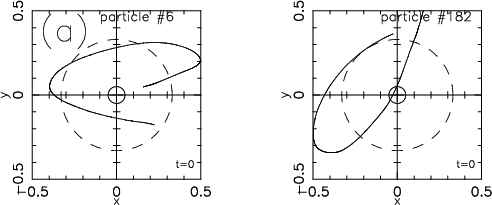}
\includegraphics[angle=0,angle=0,angle=0,height=60mm,width=160mm]{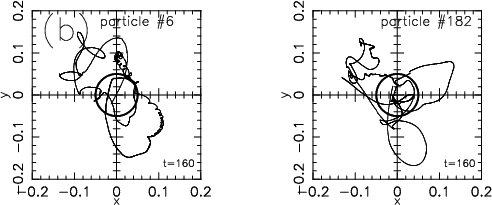}  
\caption{Two stars evolving in the core of the $N=256$ equal-mass Plummer
sphere: (a) the $x-y$ projections of particles \#6 and \#182, evolving
in the core over $4$ time units starting at $t=0$, (b) their $x-y$
projections starting at $t=160$. The outer \mybold{dashed} circle in
(a) shows the core radius $r_{\rm c}$ of the cluster at $t=0$. The
inner circle in (a) depicts the core radius at $t=160$ as shown in (b)
and illustrates the difference in scale between the two figures. }
\label{Fig:Plummer_orbits_0_cc}
\vfil }
\end{figure*}

% Caption Status: 2
%%%%%%%%%%%%%%%%%%%%%%%%%%%%%%%%%%%%%%%%%%%%%%%%%%%%%%%%%%%%%%%%%%%%%%%%%%%%%
% Figure 12: Six timeseries
%%%%%%%%%%%%%%%%%%%%%%%%%%%%%%%
\begin{figure*}
\vbox to220mm{
\vfil
\centering
\includegraphics[height=210mm,width=180mm]{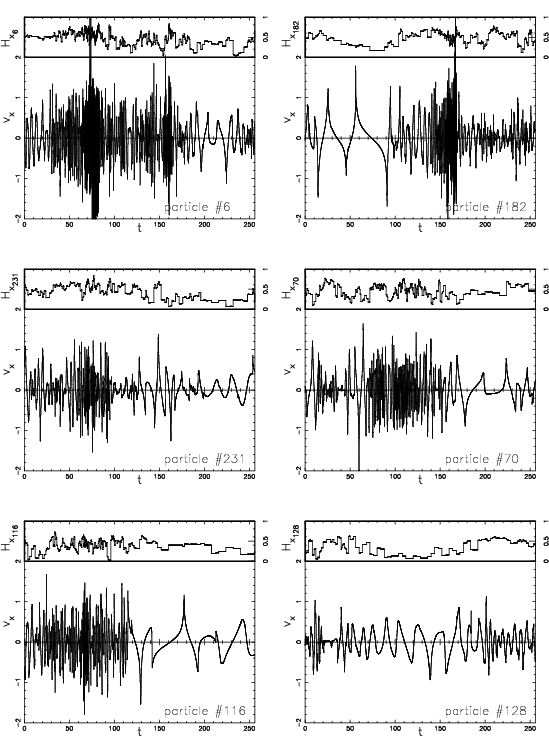}
\caption{Time evolution of $6$ individual orbits
in an $N=256$ equal-mass Plummer sphere model. Each figure shows the
complexity diagnostic obtained with respect to the $x$-coordinate of
particle $i$, and comprises the velocity time series $v_{x_{i}}$
(solid line in red, plotted against the left-hand axis) and the DWaTIM
$H_{x_{i}}$ (solid blue line in the upper inset, plotted against the
$0-1$ scale shown on the right-hand axis).}
\label{Fig:Plummer_six_timeseries}
\vfil}
\end{figure*}

%Caption Status: 2
%%%%%%%%%%%%%%%%%%%%%%%%%%%%%%%%%%%%%%%%%%%%%%%%%%%%%%%%%%%%%%%%%%%%%%%%%%%%%

\subsection{Individual orbits} 

\mybold{To illustrate the differences in orbital complexity a star may 
show between the initial time and the moment of core expansion, we
graph in Fig.\,\ref{Fig:Plummer_orbits_0_cc} the $x-y$ projections of
$2$ individual orbits during two windows of four time units, the first
running from $t=0$ to $4$ (top panels), the second running in the
interval $t=160$ to $164$ (bottom panels).}  We picked two stars that
happened to orbit within the core in each time interval (the circles
on the figure indicate $r_{\rm c}$ at the times shown). In both the
cases displayed the orbit starts off smooth and regular, but traces a
much more intricate pattern later on.

% Figure 13: DWaTIM counter N=256 Plummer sphere 
%%%%%%%%%%%%%%%%%%%%%%%%%%%%%%%%%%%% 
\begin{figure} 
\centering 
\includegraphics[height=90mm,width=85mm]{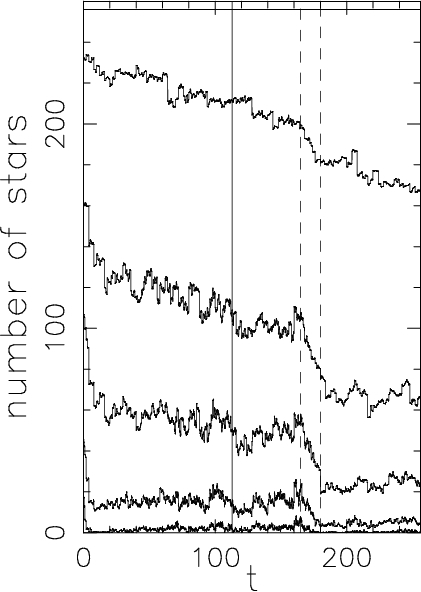} 
\caption{Cumulative number of stars of the $N=256$ 
equal-mass Plummer sphere with a DWaTIM $\Upsilon_{i}$ $(i=1,...,N)$
larger than (from top to bottom) $0.2$, $0.4$, $0.5$, $0.6$ and $0.7$,
respectively. The uppermost horizontal line shows the total number of
particles $N=256$. The solid vertical line at $t=113$ marks the
formation of the first hard binary in the system. The dashed vertical
lines indicate the phase of core expansion $(165\lesssim t\lesssim
180)$.}
\label{Fig:Plummer_dwatim_counter} 
\end{figure} 

%Caption Status: 2 
%%%%%%%%%%%%%%%%%%%%%%%%%%%%%%%%%%%%%%%%%%%%%%%%%%%%%%%%%%%%%%%%%%%%%% %%%%%%% 

These trends can be identified in a graph of the DWaTIM for these and
four other orbits as displayed on
Fig.\,\ref{Fig:Plummer_six_timeseries}.  The figure shows $v_x$ vs
time in the main frames and the DWaTIM as the top inset frame for each
case. Note the change of scales: the DWaTIM is plotted against the
scale shown at the right-hand side of each figure. Focusing on the top
two panels on Fig.\,\ref{Fig:Plummer_six_timeseries}, we can identify
the more complex phases around $160\lesssim t\lesssim 164$ of the
orbits displayed on Fig.\,\ref{Fig:Plummer_orbits_0_cc}(b) as local
peaks in the DWaTIM during this time interval. It is clear that stars
set on regular orbits initially can show more complex behavior at
later times. The opposite is also possible, as we illustrate in Fig.\,
\ref{Fig:Plummer_six_timeseries} with four more orbits also orbiting
in and out of the core region. These and many others not shown here
are typical of the wide variety of DWaTIM spectra: some orbits show
rapid fluctuations in $v_{x}$ and yield a rather broad band of base
frequencies (especially particles \#$6$ and \#$182 $). Other
trajectories have a narrower spectrum of frequencies (e.g., particle
\#$128$).

\subsection{Global behavior of the Plummer sphere}

% Figure 14: DWT N=256 Plummer sphere
%%%%%%%%%%%%%%%%%%%%%%%%%%%%%%%%%%%%
\begin{figure}
\centering
\includegraphics[height=130mm,width=80mm]{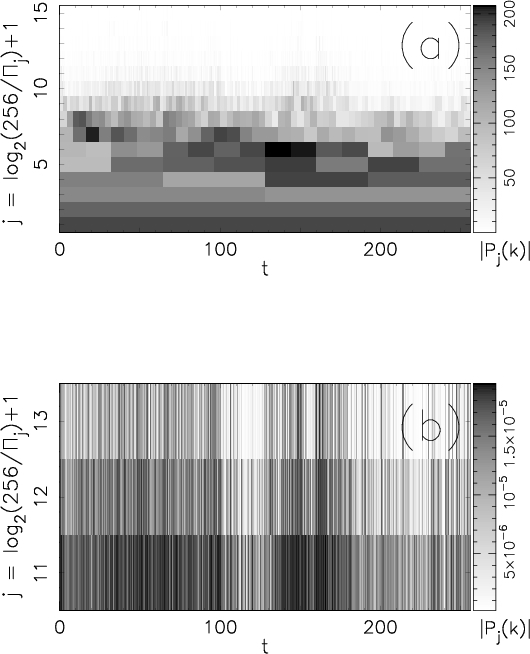}
\caption{Average discrete wavelet transform (DWaT) of the $N=256$ Plummer sphere as a function of 
time. The averaging is performed for a total of $768$ time series,
i.e. over all particles and over the three velocity components. The
figure illustrates the average amplitude of base frequency
$1/\Pi_{j}$: (a) the complete DWaT scalogram (scales $j=1,...,15$);
(b) zoom on the high-frequency scales $j=11,\,12,\rm{\,and}\,\,13$.}
\label{Fig:Plummer_dwt}
\end{figure}

\begin{figure}
\centering
\includegraphics[height=90mm,width=85mm]{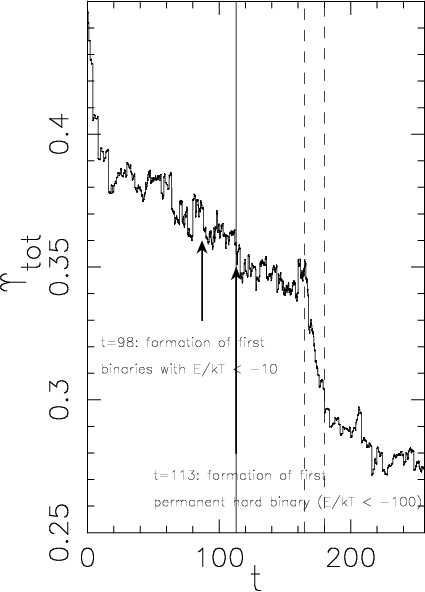}
\caption{Overall DWaTIM complexity $\Upsilon_{\rm tot}$ for an  $N=256$ equal-mass Plummer
sphere. The curve is the average of $\Upsilon_i$ over all $N$
particles and over the three velocity time series in each case (the
$x-$, $y-$ and $z-$ components). The solid vertical line at $t=113$
together with the right-most arrow mark the formation of the first
hard (stable) $E/kT < -100$ binary in the system. The left-most arrow
marks the time when the first soft (unstable) $E/kT < -10$ binary
formed. The dashed vertical lines between $165\lesssim t\lesssim 180$
bracket the period of expansion of the inner volume.  }
\label{Fig:Plummer_dwatim_wscm}
\end{figure}

%%%%%%%%%%%%%%%%%%%%%%%%%%%%%%%%%%%%%%%%%%%%%%%%%%%%%%%%%%%%%%%%%%%%%%%%%%%%%
 Fig.\,\ref{Fig:Plummer_dwatim_counter} shows as a function of time
 the cumulative number of stars whose DWaTIM $\Upsilon_{i}$
 $(i=1,...,N)$ exceeds a given threshold. The uppermost horizontal
 line on the figure denotes the total number of stars. The broken
 curves are for (from top to bottom) $\Upsilon_{i} > 0.2,\, 0.4,\,
 0.5,\, 0.6$ and $0.7$.  Once more, the vertical solid and dashed
 lines indicate the time when a hard binary first formed and the
 interval of post-collapse core expansion, respectively (cf. also
 Fig.\,\ref{Fig:Plummer_lagrad}). Broadly speaking, the orbital
 complexity decreases on the mean with time. The uppermost lines
 showing $\Upsilon_{i} > 0.2$ and $\Upsilon_{i}> 0.4$ decrease
 monotonically save for small localised fluctuations. In all the cases
 displayed, a pronounced drop in complexity is seen throughout the
 phase of core expansion, at times $160 \lesssim t \lesssim 185$. For
 instance, at the beginning of the expansion, about $110$ stars have
 $\Upsilon_{i}>0.4$ whereas at $t\approx 180$ only $75$ stars reach
 that level. The situation is similar for higher-threshold curves.
 After the formation of a hard binary but prior to post-collapse
 expansion, i.e. in the interval $113 \lesssim t \lesssim165$, the
 complexity levels off or increases slightly with time.  For example,
 the number of stars with $\Upsilon_{i} > 0.5$ increases from about
 $40$ to approximately $55$ stars in that interval. Hence, orbits that
 are already relatively complex at the time of binary formation yield
 a DWaTIM of even larger amplitude up to core-collapse and the on-set
 of the expansion phase. This trend is also found in the curve
 $\Upsilon_{i} > 0.6$ \mybold{and to a lesser extent in the
 $\Upsilon_{i} > 0.7$ curve. It may be important to note that the
 global results of Fig.\,\ref{Fig:Plummer_dwatim_counter} are not in
 contradiction with the apparent trend of increasing complexity
 depicted by the two stars of
 Fig.\,\ref{Fig:Plummer_orbits_0_cc}. Stars \#6 and \#182 of that
 figure are both part of the $8$ stars that happen to reside in the
 core around $t\approx 165$. At these times the stars have a high
 DWaTIM with values of $\Upsilon_{6}$ and $\Upsilon_{182}$ $>0.6$ (see
 also the two upper panels of
 Fig.\,\ref{Fig:Plummer_six_timeseries}). Stars \#6 and \#182
 therefore contribute to the local peak of the $\Upsilon_{i}>0.6$
 curve observed on Fig.\,\ref{Fig:Plummer_dwatim_counter} around
 $t\approx 160$. The snapshots given in
 Fig.\,\ref{Fig:Plummer_orbits_0_cc}(b) do not reflect a progressive
 increase in orbital complexity of the Plummer sphere between $0<t\leq
 164$.} The enhanced two-body scatter attributable to the newly formed
 binary can be directly measured by the DWaT.  In conclusion, both
 that event and the on-set of post-collapse expansion can be singled
 out on Fig.\,\ref{Fig:Plummer_dwatim_counter} as a local minimum and
 a local maximum, respectively, in runs of the cumulative
 $\Upsilon_{i}$ (as examplified, for instance, by the curve of
 $\Upsilon_{i}> 0.5$).\\
\indent A scalogram of the DWaT averaged over all $N=256$ Plummer sphere
particles is shown on Fig.\,\ref{Fig:Plummer_dwt}. The dark shade
illustrates base frequencies $1/\Pi_{j}$ of high amplitude, white
means zero amplitude (see also \S\ref{Subsect:pyth} and
\S\ref{Subsect:caledonian}).  
Figure \ref{Fig:Plummer_dwt}(a) brushes a global picture for all
scales $j=1,...,15$, whereas Fig.\,\ref{Fig:Plummer_dwt}(b) depicts
only the high-frequency scales $j=11,\,12,$ and $13$. Some modes are
growing in intensity and are then fading away \mybold{after}
$t \simeq 165$ units when the core starts to expand. A particularly
good example is the scale $j = 6$ which reaches progressively higher
amplitude in the interval $0 \lesssim t \lesssim 160$ before fading
away during the expansion phase of the inner volume of the sphere. The
onset of expansion triggers high-frequency modes
(cf. Fig.\,\ref{Fig:Plummer_dwt}[b]) which reflect the evolution
toward more anisotropic radial orbits.  Radial anisotropy implies more
eccentric motion relative to the centre of mass and an enhanced
spectrum of frequencies, cf. Fig.\,\ref{Fig:binaries}.\\
\indent An illustration of
the global measure of complexity $\Upsilon{\rm tot} = \sum_i \Upsilon_i /
N $ is shown on Fig.\,\ref{Fig:Plummer_dwatim_wscm}. The solid and
dashed vertical lines denote the same transitional phenomena as in all
preceding figures. On the whole, the curve of the DWaTIM depicts the same
global decrease in complexity as observed on
Fig.\,\ref{Fig:Plummer_dwatim_counter}. A close inspection of
Fig.\,\ref{Fig:Plummer_dwatim_wscm} suggests three different regimes
in the evolution of the DWaTIM indicator $\Upsilon_{\rm tot}$:

- Regime $1$ takes place between $0 < t 
\lesssim 165$. In the interval $0\leq t\lesssim 20$, we observe a  rapid drop in $\Upsilon_{\rm tot}$ 
which has no clear origin in a  physical phenomenon (binary formation, 
core-collapse, etc). The equilibrium cluster has a \mybold{dynamical time $t_{\rm d} 
\simeq 3.5$} and hence the elapsed time $t = 20 $ is $\approx 8 t_{\rm d}$ 
or eight orbital times, a further illustration of border effects
arising from a too-short time of integration. For $t \gtrsim 20$,
\mybold{the curve decreases on average until $t \simeq 110$ when a
slower but constant decline sets in. The DWaTIM $\Upsilon_{\rm tot}$
decreases from $\approx 0.38$ at $t\simeq 20$ to
$\Upsilon{\rm tot}\approx 0.35$ at $t \simeq 165$}; the formation of
soft- and hard-binaries therefore has little impact on the \textit
{global} complexity of the system. That being said, it is difficult to
disentangle the apparent trend of a drop in complexity during regime
$1$ because, first of all, the rapid drop early on is attributable to
border effects; and second, the time scale over which the trend
becomes significant is comparable to the two-body relaxation time for
that system. The data on Fig.\,\ref{Fig:Plummer_dwatim_wscm} may be
best understood in the light of
Fig.\,\ref{Fig:Plummer_dwatim_counter}. In that figure stars of low
complexity show a decrease of $\Upsilon_{i}$ between $0 < t\lesssim
110$ (see the two upper curves $\Upsilon_{i}>0.2$ and $>0.4$) whereas
the number of stars on orbits giving a diagnostics of high-complexity
remains approximately constant. This argues against border effects
reaching beyond $t \approx 20$.  The constant evolution towards more
orbits of low-complexity diagnostics drive the trend of $\Upsilon{\rm
tot}$ observed on Fig.\,\ref{Fig:Plummer_dwatim_wscm}, a statement
that the system suffers from collisional effects when stars are
shifted to higher-energy long-period orbits by two-body collisions.
Inspection of the rapid evolution of the core-radius comforts this
view.

- Regime $2$ begins around $t \simeq 165$ when the system enters
post-collapse and the central volume expands systematically. This
triggers a sudden drop in the DWaTIM from $\approx 0.35$ to $\approx
0.27$ at $t\simeq 180$.

- Regime $3$ begins (loosely speaking) at $t \simeq 180$ units when
the run of $\Upsilon_{\rm tot}$ resumes a slow decrease on the average.
This contrast in DWaTIM orbital complexity between the pre- and the
post-collapse evolution has been seen in a set of ten $N=256$ test
simulations we performed with different random seeds. An attempt to
obtain ensemble averaged results for this set of simulations proved
fruitless owing to large scatter e.g., in the core-collapse time
between two individual runs.

% Figure 15: DWaTIM N=256 Plummer sphere
%%%%%%%%%%%%%%%%%%%%%%%%%%%%%%%%%%%%
%\begin{figure*}
%\vbox to140mm{
%\vfil
%\centering
%\includegraphics[height=110mm,width=160mm]{revised_images/dwatim_plummer_mended.jpg}
%\caption{Overall DWaTIM complexity $\Upsilon_{\rm tot}$ of the $N=256$ equal-mass Plummer
%sphere. The curve has been obtained by averaging over all $N$
%particles of the system and over the 3 components $x$, $y$ and
%$z$. The dotted vertical line at $t=64$ delimits the region where the
%DWaTIM may become unreliable due to border effects (see \S 2.1.2). The
%solid vertical line at $t\simeq 113$ marks the changeover from the
%pre-collapse phase to the post-collapse phase (see also
%Fig.\,\ref{Fig:Plummer_lagrad}). The dashed vertical lines between
%$165\lesssim t\lesssim 180$ indicate the onset of a notable core
%expansion. The occurrence of core-collapse and of core expansion
%yields an overall decrease of orbital complexity. The arrows mark the
%instant of first soft binary formation with $E/kT < -10$ and the
%instant of formation of the first hard binary with $E/kT < -100$,
%respectively.}
%\label{Fig:Plummer_dwatim_wscm}
%\vfil}
%\end{figure*}
%%%%%%%%%%%%%%%%%%%%%%%%%%%%%%%%%%%%%%%%%%%%%%%%%%%%%%%%%%%%%%%%%%%%%%%%%%%%%%
%%%%%%%%%%%%%%%%%%%%%%%%%%%%%%%%%%%%%%%%%%%%%%%%%%%%%%%%%%%%%%%%%%%%%%%%%%%%%
% Figure 15: DWaTIM  N=256 Plummer sphere
%%%%%%%%%%%%%%%%%%%%%%%%%%%%%%%%%%%%

%Caption Status: 2
%%%%%%%%%%%%%%%%%%%%%%%%%%%%%%%%%%%%%%%%%%%%%%%%%%%%%%%%%%%%%%%%%%%%%%%%%%%%%

\section{Discussion}

%The close connection between information entropy and the
%concept of entropy used in thermodynamics is described by several
%authors (e.g.,
%\citealt{Brillouin53}; \citealt{Petersen83}). 

% one could discuss the possible correlation between complexity and Lyapunov exponents

We presented a method to compute the time-dependent orbital complexity
in $N$-body simulations. The gravitational $N$-body problem is
described by the $3N$ second order ordinary differential equations of
Eq.\,(\ref{Eqn:grav}). We extract a discrete wavelet transform
information measure (DWaTIM) from the velocity time series of the
individual particles of the $N$-body simulation. We apply the technique
to several few-body problems and to a larger $N=256$ particle
simulation. The method captures the time-dependent changes in the
dynamics of three- and five-body systems and furthermore quantifies
orbital complexity continuously in time. For example, we recovered and
quantified the dynamically more complex phases of the well-studied
Pythagorean problem (see \S\ref{Subsect:pyth}) as well as the complex
dynamics of a perturbed Caledonian configuration (see
\S \ref{Subsect:caledonian}).  We also applied the method to a set of
$N=256$ equal-mass Plummer spheres (see \S \ref{Sect:plummer}). We
found that, on a global scale, orbital complexity decreases during the
evolution. The occurrence of core-collapse and the subsequent core
expansion causes a considerable drop in overall DWaTIM complexity
$\Upsilon_{\rm tot}$. Furthermore, we observed that the complexity of
individual orbits with a DWaTIM $\Upsilon_{i}\geq 0.5\, (i=1,...,N)$
at the instant of core-collapse tends to increase until the occurrence
of core expansion.\\
\indent We opted for a DWaTIM implementation that allowed 1) to identify qualitative
changes in orbital dynamics in a quasi-instantaneous manner and 2) to
provide a continuous quantification of time-dependent complexity on a
well-defined gauge between $0$ and $1$. One should however bear in
mind that an
\textit{absolute} measure of complexity of a dynamical system cannot
be obtained by the method. Complexity ultimately depends on the range
of scales over which the system is studied. The DWaTIM is a
band-limited measure and therefore strongly depends on the resolution
of the time series that is subjected to analysis. Experience drawn
from several test cases shows that a repeat calculation with truncated
frequency range is desirable to confirm the convergence of the
diagnostics.\\
\indent Border effects are an unwanted artifact
of the wavelet transform. In order to reduce the bias so introduced we opted to implement a
\textit{discrete} wavelet transform and to select cubic spline functions as
mother wavelets.  This DWaT implementation avoids redundant wavelet
coefficients. This helps in obtaining a proper measure of complexity
for the two limiting cases of a unique base frequency sinusoid (DWaTIM
= 0) and of a white noise signal (DWaTIM =~1), and thus to recover a
well-defined gauge of complexity. We obtain reliable diagnostics of
complexity whenever the orbit is integrated over a minimum of $8$
complete oscillations (see
\S\ref{Subsect:pyth}).

The method is computationally inexpensive. For example, the analysis
of a time series of length $Q=2^{13}$ data points takes approximately
$1$ second on a Pentium IV 2.4 GHz workstation with 1.2 GB RAM
memory. It is then possible to obtain a complexity diagnostic of an
$N=10^5$ body system in about $80$ processor-hours. A parallel
implementation of the scheme is currently under development.

\section*{Acknowledgments}
\mybold{We would like to thank the referee, Daniel Carpintero, for a
critical reading of an earlier version of this manuscript and for his
helpful comments.} The work was supported by grant BFR-04/55 of the
Ministry for Higher Education and Research, Grand-Duchy of Luxembourg
and by NWO under grant number 643.200.503. Additional support was
provided by the European Doctoral College (EDC) in Strasbourg, France,
by LKBF and the Netherlands Advanced School for Astrophysics (NOVA).

%\label{}
% \appendix

% \section{}
% \label{}

% Bibliographic references with the natbib package:
% Parenthetical: \citep{Bai92} produces (Bailyn 1992).
% Textual: \citet{Bai95} produces Bailyn et al. (1995).
% An affix and part of a reference:
%   \citep[e.g.][Ch. 2]{Bar76}
%   produces (e.g. Barnes et al. 1976, Ch. 2).

%\newpage

\end{document}